\documentclass{egpubl}
 
\JournalSubmission    
\usepackage[T1]{fontenc}
\usepackage{dfadobe}  

\usepackage{booktabs} 

\usepackage{tabu}               
\usepackage{multirow}
\usepackage{adjustbox}

\newcolumntype{P}[1]{>{\centering\arraybackslash}p{#1}}

\newcommand{\subsmallfigsize}{0.126}
\newcommand{\subfigfoursize}{0.147}
\newcommand{\subfigsbmcsize}{0.129}
\newcommand{\subfigonefigsize}{0.1285}
\newcommand{\subfigfailuresize}{0.157}
\newcommand{\etal}{\textit{et al. }}
\newcommand{\old}[1]{}

\biberVersion
\BibtexOrBiblatex
\usepackage[backend=biber,bibstyle=EG,citestyle=alphabetic,backref=true]{biblatex} 
\electronicVersion
\PrintedOrElectronic

\usepackage{egweblnk} 

\title[Auxiliary Features-Guided Super Resolution for Monte Carlo Rendering]{Auxiliary Features-Guided Super Resolution for Monte Carlo Rendering}

\author[Qiqi Hou]
{\parbox{\textwidth}{\centering Qiqi Hou\thanks{Email: qiqi.hou2012@gmail.com}\orcid{0009-0009-3472-6401}
        and Feng Liu\thanks{Email: fliu@pdx.edu}\orcid{0000-0002-5399-6214} 
        }
        \\
{\parbox{\textwidth}{\centering Portland State University, USA
       }
}
}

\begin{document}

\teaser{
\def\figoneindent{-4.mm}
\vspace{-0.4in}
        \begin{tabular}{cc}
        \hspace{-5mm}
        \begin{adjustbox}{valign=t}
        \scriptsize
            \begin{tabular}{c}
              \includegraphics[width=0.336\textwidth]{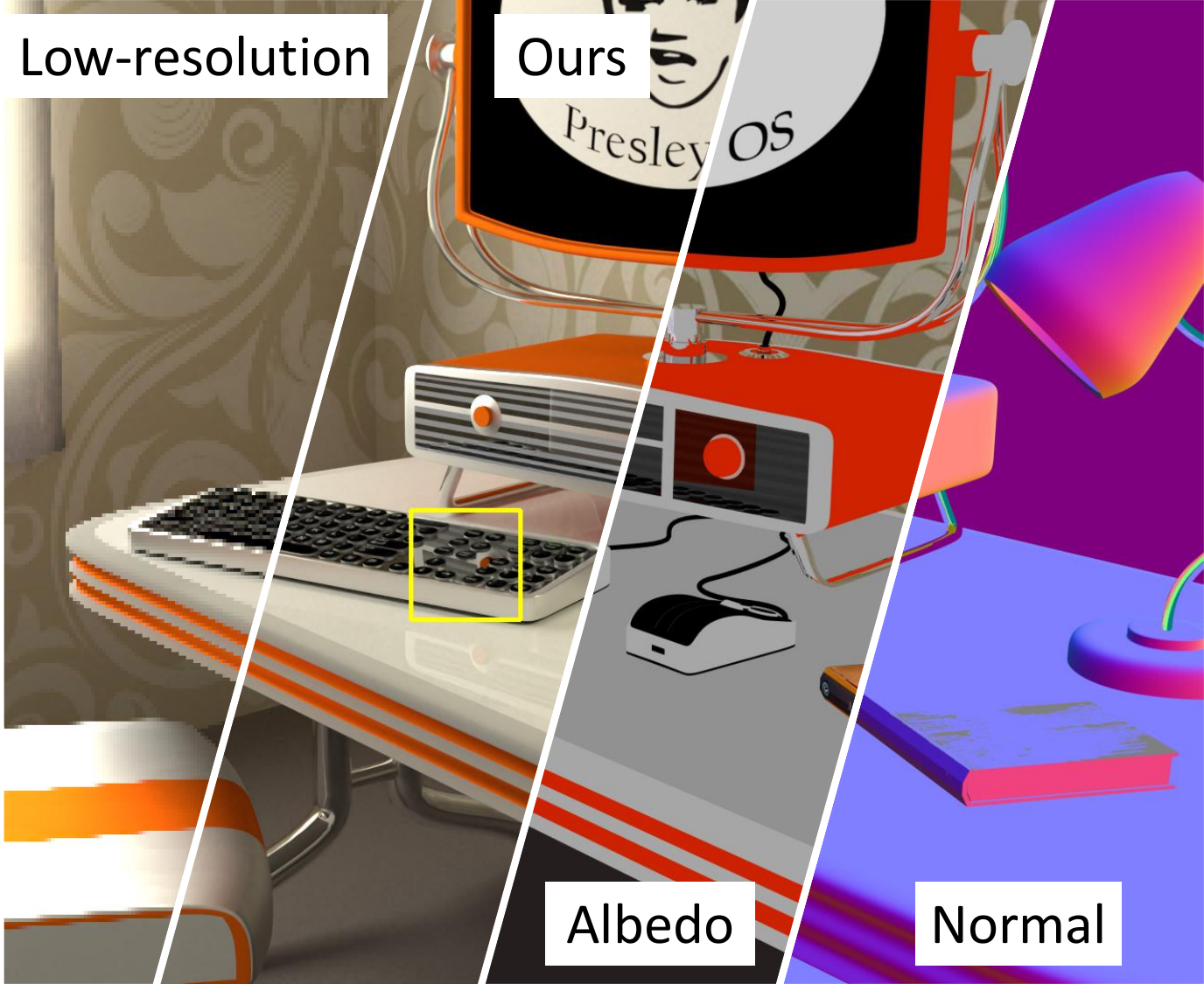}
                \\
                Ground Truth
            \end{tabular}
        \end{adjustbox}
        \hspace{-8.2mm}
        &
        \begin{adjustbox}{valign=t}
        \scriptsize
            \begin{tabular}{ccccc}
                \includegraphics[width=\subfigonefigsize\textwidth]{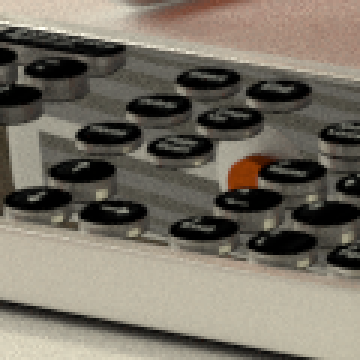} \hspace{\figoneindent} &
                \includegraphics[width=\subfigonefigsize\textwidth]{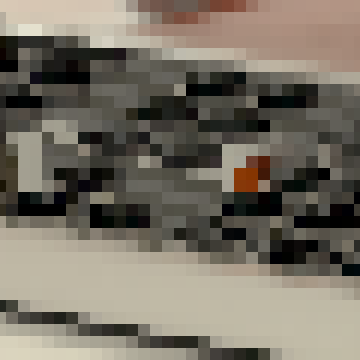} \hspace{\figoneindent} &
                \includegraphics[width=\subfigonefigsize\textwidth]{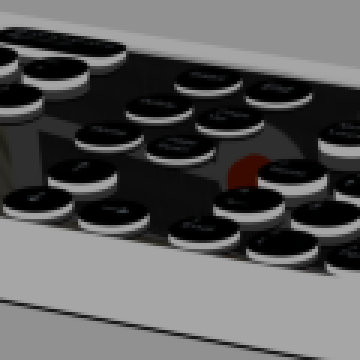} \hspace{\figoneindent} &
                \includegraphics[width=\subfigonefigsize\textwidth]{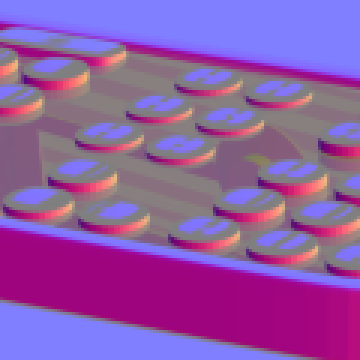} \hspace{\figoneindent} &
                \includegraphics[width=\subfigonefigsize\textwidth]{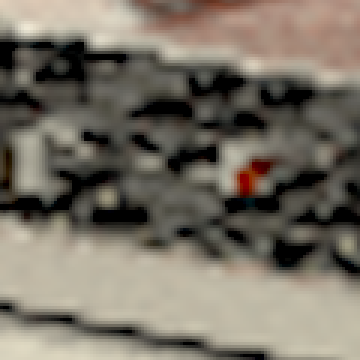}
                \\
                Ground Truth \hspace{\figoneindent} &
                LR \hspace{\figoneindent} &
                Albedo \hspace{\figoneindent} &
                Normal \hspace{\figoneindent} &
                Bicubic
                \\
                \includegraphics[width=\subfigonefigsize\textwidth]{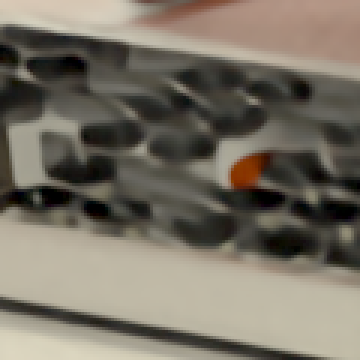} \hspace{\figoneindent} &
                \includegraphics[width=\subfigonefigsize\textwidth]{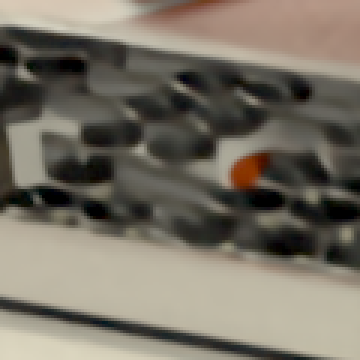} \hspace{\figoneindent} &
                \includegraphics[width=\subfigonefigsize\textwidth]{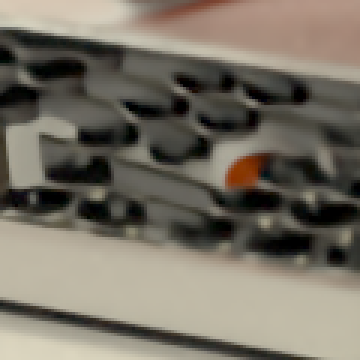} \hspace{\figoneindent} &
                \includegraphics[width=\subfigonefigsize\textwidth]{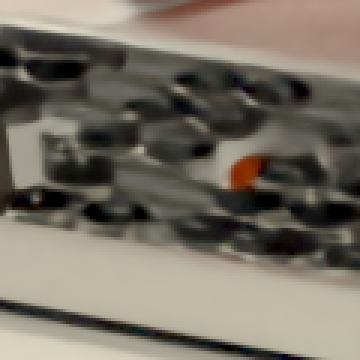} \hspace{\figoneindent} &
                \includegraphics[width=\subfigonefigsize\textwidth]{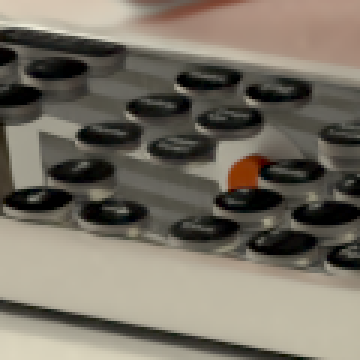}
                \\ 
                EDSR~\cite{lim2017enhanced} \hspace{\figoneindent} &
                RCAN~\cite{zhang2018image} \hspace{\figoneindent} &
                SwinIR~\cite{liang2021swinir} \hspace{\figoneindent} &
                MSSPL~\cite{hou2021fast} \hspace{\figoneindent} &
                Ours $\times4$
                \\
            \end{tabular}
        \end{adjustbox} 
        \\
    \end{tabular}\vspace{-0.0in}
    \caption{Our method uses fast-to-compute high-resolution auxiliary features to support super resolution of Monte Carlo rendering results. Our method generates high-quality visual details than both off-the-shelf super resolution methods and MSSPL, a dedicated super resolution method for Monte Carlo renderings that uses both more high-resolution auxiliary features and the corresponding high resolution low sample rendering result.  
    }
    \label{fig:fig1}
    \vspace{-0.0in}
}

\maketitle
\begin{abstract}
This paper investigates super resolution to reduce the number of pixels to render and thus speed up Monte Carlo rendering algorithms. While great progress has been made to super resolution technologies, it is essentially an ill-posed problem and cannot recover high-frequency details in renderings. To address this problem, we exploit high-resolution auxiliary features to guide super resolution of low-resolution renderings. These high-resolution auxiliary features can be quickly rendered by a rendering engine and at the same time provide valuable high-frequency details to assist super resolution. To this end, we develop a cross-modality Transformer network that consists of an auxiliary feature branch and a low-resolution rendering branch. These two branches are designed to fuse high-resolution auxiliary features with the corresponding low-resolution rendering. Furthermore, we design residual densely-connected Swin Transformer groups to learn to extract representative features to enable high-quality super-resolution. Our experiments show that our auxiliary features-guided super-resolution method outperforms both super-resolution methods and Monte Carlo denoising methods in producing high-quality renderings.


\end{abstract}  

 \vspace{-0.in}
\section{Introduction}
\label{sec:intro}
Monte Carlo rendering algorithms are now widely used to generate photo realistic computer graphics images for applications such as visual effects, video games, and computer animations. These algorithms generate a pixel's color by integrating over all the light paths arriving at a single point~\cite{cook1984distributed}. To rendering a high-quality image, a large number of rays need to be cast for each pixel, which makes Monte Carol rendering a slow process.

\old{The ray tracing method is one of the most successful physical photo-realistic image render methods and has attracted significant attention. Due to its superior capacity in generating high-quality computer graphic images, it has been widely adopted by commercial renderers. It plays a vital role in visual effects, video games, and video editing. Fundamentally, the ray tracing method integrated over all the light paths arriving at a single point and bounced back to a light source~\cite{cook1984distributed}. However, ray tracing methods need to cast numerous rays and often take minutes or hours for rendering to obtain visually satisfactory results without noticeable noise. The immense computational requirement of path tracing methods poses a challenge in mapping the path tracing method on devices with limited computing resources.}

A great amount of effort has been devoted to speeding up Monte Carlo rendering. The core idea is to reduce the number of rays for each pixel. For instance, numerous denoising algorithms are now available to reconstruct a high-quality image from a rendering produced at a low sampling rate. Such Monte Carlo denoising algorithms often use auxiliary features generated by a rendering algorithm to help denoise the noisy rendering result. The recent deep neural network-based denoising algorithms can now generate very high-quality images at a fairly low sampling rate~\cite{bako2017kernel,chaitanya2017interactive,kuznetsov2018deep,gharbi2019sample}.

Monte Carlo rendering can also be sped up by reducing the number of pixels to render. For example, pixels from the frames that have already been rendered can be warped to generate frames in-between existing frames to increase the frame rate~\cite{briedis2021neural} or to generate future frames to reduce latency~\cite{guo2021extranet}. Another approach is to only render one pixel for a block of neighboring pixels to further reduce the total number of pixels to render. This can be implemented by first rendering a low-resolution image and then applying super resolution to increase its resolution~\cite{xiao2020neural,hou2021fast}. As super resolution is a fundamentally ill-posed problem, it alone often cannot recover high-frequency details from only the low-resolution rendering. To address this problem, Hou~\etal render a high-resolution rendering with a low sampling rate and use that together with the high-resolution auxiliary features to help super resolve the low-resolution rendering rendered at a high sampling rate. While this method produces a high-quality result, it needs to render the high-resolution image at a low sampling rate, which still takes a considerable amount of time~\cite{hou2021fast}.

\emph{Can we only use the fast-to-obtain high-resolution auxiliary features without the high-resolution-low-sample rendering to effectively assist super resolution of the corresponding low-resolution rendering?} If so, we can further speed up Monte Carlo rendering. We are encouraged by the recent work on neural frame synthesis that showed fast-to-obtain auxiliary features of the target frames can greatly help interpolate or extrapolate the target frames~\cite{briedis2021neural,guo2021extranet}. On the other hand, Hou~\etal showed that using a wide range of auxiliary features and the high-resolution-low-sample rendering help super resolution more than only using a subset of auxiliary features within their own deep neural network-based super resolution framework~\cite{hou2021fast}. Therefore, if we only use a small number of fast-to-compute auxiliary features, we need to have a better super resolution method.     

This paper presents a Cross-modality Residual Densely-connected Swin Transformer (XRDS) for super resolution of a Monte Carlo rendering guided by its auxiliary features. For the seek of  speed, we only use two auxiliary features: albedo and normal. To effectively use these features, we design a super resolution network based on Swin Transformer that recently has been shown powerful for a wide variety of computer vision tasks. Our Transformer network has two branches, one for the low resolution rendering and the other for the auxiliary features. Such two branches are designed to perform cross-modality fusion to effectively use auxiliary features to assist super resolution of the low-resolution rendering. While the auxiliary feature branch consists of convolutional blocks, the branch for the low-resolution rendering consists of a sequence of residual densely-connected Swin Transformer blocks to extract effective features. The features from the two branches are combined together using a cross-modality fusion module and are finally used to generate the high-resolution high-quality rendering. 

This paper contributes to Monte Carlo rendering as follows. First, we present the first super resolution approach to Monte Carlo rendering that only uses fast-to-compute high-resolution auxiliary features to enable high-quality upsampling of a low-resolution rendering. Second, we design a dedicated Cross-modality Swin Transformer-based super resolution network that can learn to effectively combine high-resolution auxiliary features with the corresponding low-resolution rendering to generate the final high-resolution high-quality image. Third, our experiments show that our method outperforms super-resolution and denoising methods in producing high-quality renderings.

\section{Related Work}
\label{sec:related} 
\begin{figure*}[ht]
\footnotesize
  \includegraphics[width=1.0\textwidth]{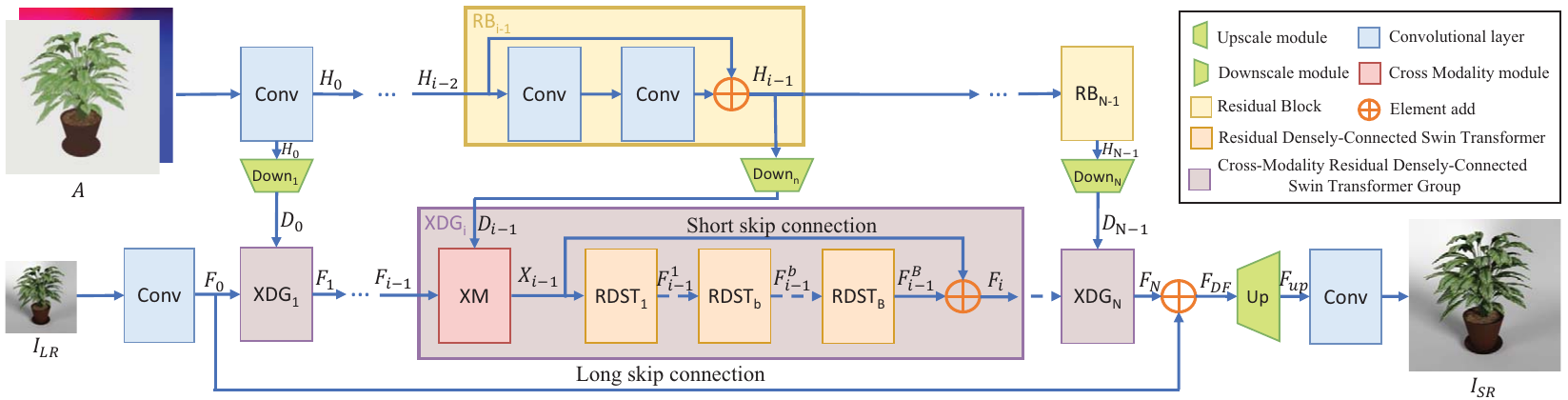}\vspace{-0.1in}
  \caption{The architecture of our network. Our network takes a low-resolution rendering  and its corresponding fast-to-compute high-resolution auxiliary features as input and predicts the final high-resolution-high-quality image.}
  \label{fig:network}\vspace{-0.10in}
\end{figure*}

This section briefly discusses relevant work to our paper, including Monte Carlo denoising, super resolution, and vision Transformers. 

\noindent\textbf{Monte Carlo Denoising.} Monte Carlo rendering algorithms need numerous samples per pixel to generate a high-quality rendering~\cite{cook1984distributed,kajiya1986rendering}. With insufficient samples, the rendering results suffer from noise. To address this problem, many Monte Carlo denoising methods have been developed to  reconstruct high-quality renderings from only a small number of samples. Traditional methods reconstruct renderings in a similar way to general image denoising methods by designing specific denoising kernels based on image variance or geometric features or directly regress the final result~\cite{dammertz2010edge,jensen1995optimizing,laine2007incremental,lee1990note,mccool1999anisotropic,rushmeier1994energy,segovia2006non,sen2012filtering,xu2005novel}. Zwicker~\etal provided a good survey on this~\cite{zwicker2015recent}. Alternatively, adaptive sampling algorithms can be used to reduce the overall sample numbers for the whole image~\cite{bolin1998perceptually,egan2009frequency,jensen2001realistic,li2012sure,meyer2006statistical,moon2016adaptive,overbeck2009adaptive,rousselle2011adaptive,walter2006multidimensional,ward1988ray}. 

Recently, deep neural network-based denoising methods have shown impressive performance. These methods learn to reconstruct high-quality renderings from small number of samples. In their seminal work, Kalantari~\etal estimated optimal filter parameters using a multi-layer perceptron neural network~\cite{kalantari2015machine}. Bako~\etal estimated spatially adaptive kernels for denoising in a convolutional manner~\cite{bako2017kernel}. Vogels~\etal extended the concept of kernel prediction methods to temporal denoising~\cite{vogels2018denoising}. With asymmetric loss functions, their method could produce high-quality results for a sequence of frames. Chaitanya~\etal developed a recurrent autoencoder to denoise a sequence of frames while maintaining temporal stability~\cite{chaitanya2017interactive}. Xu~\etal developed an adversarial approach to Monte Carlo rendering denoising that can greatly reduce artifacts such as blurs and unfaithful details from denoising results~\cite{xu2019adversarial}. Gharbi~\etal developed a kernel splatting network that reconstructs the final image by splatting samples to pixels according to the estimated splatting kernels~\cite{gharbi2019sample}. Munkberg~\etal proposed to filter auxiliary layers of individual samples~\cite{munkberg2020neural}. Their method works well on outliers and complex visibility. Hasselgren~\etal proposed a neural spatial-temporal sampling method for Monte Carlo video denoising~\cite{hasselgren2020neural}. Their method first estimates the sampling map from the temporal reprojection and auxiliary features and then denoises the resulting image generated using the sampling map to produce high-quality results.  
Zheng~\etal proposed an ensemble denoising technique that learns to combine multiple denoiser together~\cite{zheng2021ensemble}. Yu~\etal designed a transformer-based neural network for Monte Carlo denoising~\cite{yu2021monte}. Their network consists multi-scale feature extractor and a self-attention module and achieved promising results.  Unlike these denoising methods, our method explores an orthogonal approach that speeds up Monte Carlo rendering by reducing the number of pixels to render via super-resolution.








\vspace{0.1in}\noindent\textbf{Super resolution.} Super resolution is a classic problem in computer vision. It aims to reconstruct a high-resolution image from the low-resolution input. Recently, the state of the art of super resolution research has been advanced significantly due to the use of deep neural networks~\cite{dong2014learning,ahn2018fast,haris2019recurrent,hui2018fast,kim2016deeply,lim2017enhanced,li20193d,liu2019hierarchical,xu2019towards,zhang2018learning,zhang2019image}. Specifically, in their seminal work, Dong~\etal trained a three-layer fully convolutional neural network for single image super resolution~\cite{dong2014learning}. Since that, a large variety of super resolution deep neural networks have been invented by leveraging carefully designed network architectures, including residual blocks~\cite{he2016deep}, densely connected blocks~\cite{huang2017densely,zhang2018residual}, channel attention blocks~\cite{hu2018squeeze,zhang2018image}, transformers~\cite{liang2021swinir}, and others. 

Super resolution has recently been explored to speed up rendering. Xiao~\etal designed a super resolution network that takes both the low-resolution rendering and the corresponding low-resolution auxiliary features as input and outputs a high-resolution frame. They leveraged neighboring frames to further improve super resolution quality~\cite{xiao2020neural}. While their method was designed for a rasterization-based renderering engine, in principal, it can be applied to Monte Carlo rendering. Thomas~\etal combined super resolution and Monte Carlo denoising for videos. Their network takes a low-resolution rendering as well as a warped previous frame as input and produces a high-resolution frame~\cite{thomas2022temporally}. However, super resolution is essentially an ill-posed problem and cannot fully recover missing high-frequency visual details from only the low-resolution input. To address this problem, Hou~\etal developed a super resolution approach based on multiple-resolution sampling. Their method first renders a low-resolution image at a high-sampling rate and a high-resolution image at a low sampling rate. Their method then exploits the high-resolution noisy image to recover high frequency visual details~\cite{hou2021fast}. While their method generates high-quality renderings, it needs to render the high resolution noisy image and auxiliary features, which takes a considerable amount of time. Different from the above methods, our method obtains high-frequency information from only fast-to-compute high-resolution auxiliary features as inspired by recent interpolation and extrapolation methods that use fast-to-compute auxiliary features of the target frames to help generate the target frames~\cite{guo2021extranet,briedis2021neural}. 

\old{Our method is also relevant to recent approaches which use cheap auxiliary feature buffers to improve rendering performance. Compared to the shading layers, these auxiliary geometry buffers are relatively easy to acquire. Guo~\etal proposed a extrapolation network for real-time rendering~\cite{guo2021extranet}. In their network, they generate the cheap auxiliary features for the next frame, and they only need to predict the shading for the next frame. Their method achieves high-quality results while only requiring several milliseconds for rendering. Briedis~\etal developed a neural frame interpolation network for rendering~\cite{briedis2021neural}. They leverage the cheap auxiliary buffers of intermediate frames to improve the network's capability for compositing and occlusion handling. Inspired by the success of the above methods, our method introduces cheap auxiliary geometry buffers to super-resolution in Monte Carlo Denoising.} 

\noindent\textbf{Vision transformer.} Transformer was initially designed for natural language process tasks~\cite{vaswani2017attention}. Due to its self-attention mechanism, it can efficiently capture the long-term information from the input. Recently, transformer networks have attracted considerable attention in the computer vision community and achieved great success in various computer vision tasks, including image recognition~\cite{liu2021swin}, object detection~\cite{carion2020end}, semantic segmentation~\cite{zheng2021rethinking}, and image restoration~\cite{liang2021swinir}. Dosovitskiy~\etal developed the first transformer network for image recognition~\cite{dosovitskiy2020image}. They split the input image into image patches and then feed these image patches as tokens to the transformer network. Chen~\etal presented an image processing transformer for various restoration problems and demonstrated that pretraining on large datasets could greatly improve the capability of a transformer network for low-level computer vision tasks~\cite{chen2020pre}. Liang~\etal developed SwinIR for image restoration. Their network adapted the Swin Transformer~\cite{liu2021swin} as their backbone and achieved promising results~\cite{liang2021swinir}. However, transformer for the Monte-Carlo denoising is less explored. Inspired by the success of these vision transformer networks, we are the first to design a dedicated cross-modality transformer network for super resolution of Monte Carlo renderings that can effectively leverage fast-to-compute high-resolution auxiliary features to recover high-frequency visual details when upsampling a low-resolution rendering.



\section{Our Method}
\label{sec:method}
\vspace{0.05in}

This paper proposes a super resolution method guided by the fast-to-compute auxiliary features to speed up the Monte Carlo rendering.  Our method takes a low resolution rendering $I_{LR}$ and its high-resolution fast-to-compute auxiliary features $A$ as input, and outputs the corresponding high-quality high-resolution result $I_{SR}$. The high-resolution auxiliary features provide the essential high-frequency information for the super-resolution. 

Different from the previous work~\cite{hou2021fast}, which leverages a wide range of auxiliary features, our method \textit{only} employs the auxiliary features that can be computed very fast~\cite{briedis2021neural}, including albedo and normal. On the one hand, although our method doesn't leverage the shading layers, albedo and normal could provide a lot of high-frequency information, e.g., the texture of the material, which is essential for super-resolution. As we will show, it can help improve the super-resolution results. On the other hand, albedo and normal can be computed fast~\cite{briedis2021neural}. It not only reduces the rendering time but also enables us to render these high-resolution layers at a relatively higher sampling rate, which typically contains fewer artifacts, such as aliasing.  

\begin{figure}[t]
\footnotesize
  \includegraphics[width=0.48\textwidth]{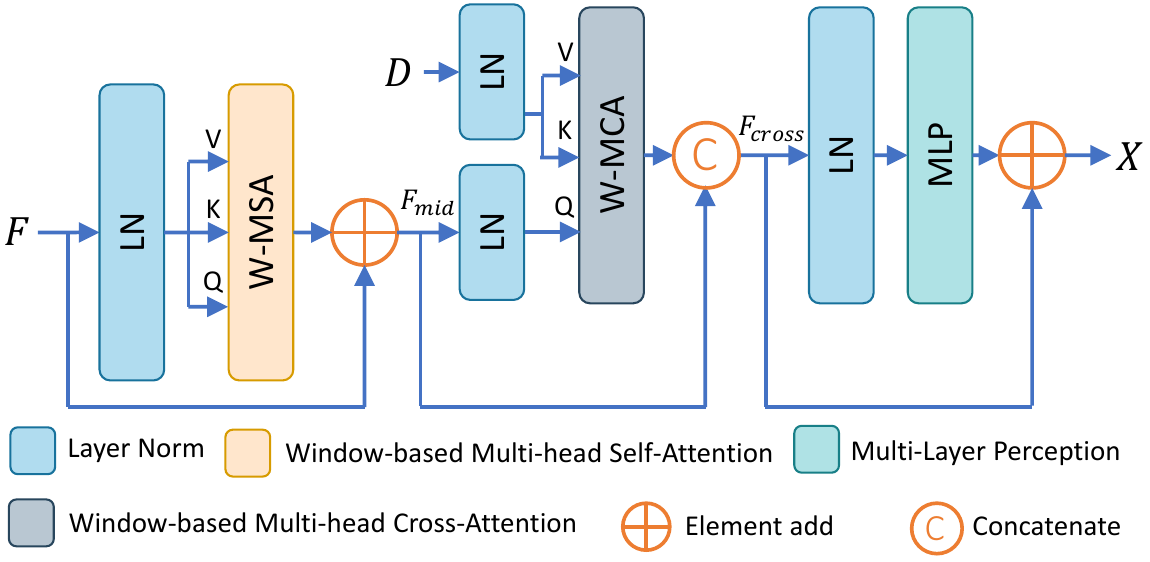}\vspace{-0.in}
  \caption{The cross-modality module. It takes feature $F$ from the low-resolution rendering branch and $D$ from the auxiliary feature branch, and outputs the fused feature $X$. }
  \label{fig:cross}\vspace{-0.15in}
\end{figure}

We design a cross-modality transformer network to effectively fuse two categories of visual input, namely the low-resolution rendering and its corresponding high-resolution auxiliary features, to recover visual details. Figure~\ref{fig:network} shows the architecture of our network. It contains two parallel branches, one for the low-resolution rendering and the other for the corresponding high-resolution auxiliary features.




\noindent\textbf{Auxiliary feature branch.} The auxiliary feature branch takes auxiliary features as inputs, which provide essential high frequency visual details. As discussed above, we select albedo and normal, which are relatively fast to acquire. Since this branch processes high-resolution input, we design a shallow architecture for the sake of memory and speed. As shown in Figure~\ref{fig:network}, we employ a convolutional layer and $N=3$ residual blocks (RB)~\cite{he2016deep} in a sequence to get the features $\{H_i\}_{i=0}^{N-1}$
\begin{equation}
    \begin{array}{llll}
        H_0 &= &f_{conv}^A(A),  \\
        H_i &= &f_{RB}^i (H_{i-1}), &\quad i = 1, \cdots, N-1,
    \end{array}
\end{equation}
where $f_{conv}(\cdot)$ indicates the convolution operation. $f_{RB}(\cdot)$ indicates the operation of a residual block. In our experiments, we set the channels as 32 for the auxiliary feature branch. 


We then obtain the downsampled features $\{D_i\}_{i=0}^{N-1}$ with a group of deshuffle layers~\cite{hou2021fast}, which is able to downscale the feature while keeping the high frequency information.
\begin{equation}
    D_i = f_{DSF}^{i+1} (H_{i}), \quad i = 0, \cdots, N-1
\end{equation}
where $f_{DSF}(\cdot)$ indicates the deshuffle layer. 


\noindent\textbf{Low resolution rendering branch.} Following  the recent works on image super resolution~\cite{zhang2018image,zhang2018learning,liang2021swinir}, we first adopt a $3\times3$ convolutional layer with 64 channels to get the shallow feature from the low resolution rendering $I_{LR}$.
\begin{equation}
    F_0 = f_{conv}^{LR}(I_{LR})
\end{equation}
We feed the resulting feature $F_0$ to a sequence of cross-modality residual densely-connected Swin Transformer groups (XDG).
\begin{equation}
    F_i = f_{XDG}^{i}(F_{i-1}, D_{i-1}), \quad i = 1, \cdots, N
\end{equation}
where $f_{XDG}(\cdot)$ indicates the XDG module. $N$ indicates the number of XDG. We choose $N = 3$ in our experiments. XDG is designed to fuse the auxiliary features $D_{i-1}$ and the low-resolution rendering features $F_{i-1}$. It consists of a cross-modality module (XM) and a sequence of residual densely-connected Swin Transformer blocks (RDST). Specifically, XM is designed to fuse the local information from the low-resolution rendering and the high frequency information from the auxiliary features, while the RDST sequence learns more dedicated representations for super resolution from them.

\noindent\textbf{Cross-modality module (XM).} Inspired by the success of Swin Transformer~\cite{liu2021swin,liang2021swinir} and Transformer Decoder~\cite{geng2022rstt}, we design XM based on Swin Transformer, which can efficiently model the long-range dependency. Figure~\ref{fig:cross} shows the architecture of XM.  It takes features $F$ from the low-resolution rendering branch and features $D$ from the auxiliary feature branch as input and outputs the fused feature $X$. It consists of Layer Norm layers (LN), a Window-based Multi-head Self-Attention layer (W-MSA), a Window-based Multi-head Cross Attention layer (W-MCA), and a Multi-Layer Perception layer (MLP). The key idea behind XM is to combine the features $F$ from the low-resolution rendering branch with the features $D$ from the high-resolution auxiliary branch using cross-attention, creating a more comprehensive representation for super resolution. The process starts by extracting intermediate features $F_{mid}$ from $F$, which serve as the "query" $Q$. From $D$, which holds high-resolution information, the "key" $K$ and "value" $V$ are extracted. Then, the cross-attention is calculated following~\cite{vaswani2017attention} and combined with $F_{mid}$ to generate $F_{cross}$. Finally, an MLP layer is used to integrate the features from the low-resolution branch and the cross-attention.

\begin{figure}[t]
\footnotesize
  \includegraphics[width=0.48\textwidth]{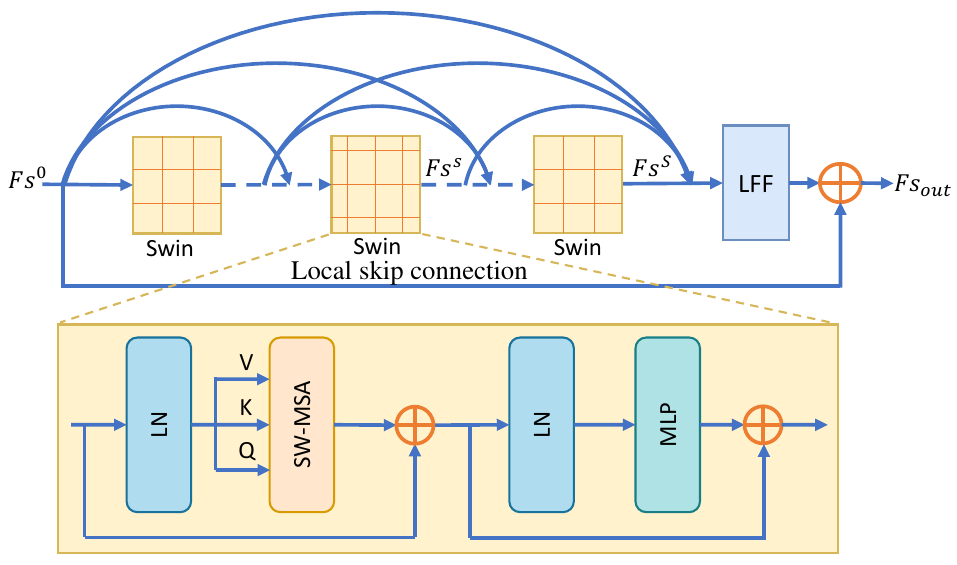}\vspace{-0.in}
  \caption{The residual densely-connected Swin Transformer block (RDST). Red lines indicate the window partitions. }
  \label{fig:rdst}\vspace{-0.3in}
\end{figure}

\begin{table}[ht]
    \centering
    \small
    \begin{tabu}{P{1.4cm}P{0.45cm}P{0.9cm}P{0.45cm}P{0.9cm}P{0.45cm}P{0.9cm}}
            \toprule
            \multirow{2}[2]{*}{Methods} & \multicolumn{2}{c}{$\times 2$} & \multicolumn{2}{c}{$\times 4$} & \multicolumn{2}{c}{$\times 8$} \\ \cmidrule(lr){2-3}  \cmidrule(lr){4-5}  \cmidrule(lr){6-7} 
             &PSNR &RelMSE &PSNR &RelMSE &PSNR &RelMSE \\ \midrule
            
             Bicubic&30.57    &0.0141   &25.39    &0.0858  &22.36    &0.2473\\
             
             EDSR &32.01    &0.0079    &30.70    &0.0119  &27.97    &0.0241\\
             RCAN &32.03   &0.0084    &30.73    &0.0117  &27.92    &0.0253  \\ 
             SwinIR &31.05 &0.0118 &30.78 &0.0152 &28.04 &0.0364\\
             MSSPL &38.40    &0.0015    &34.27    &0.0039    &31.08    &0.0079\\ \midrule
             Ours &\textbf{42.48} &\textbf{0.0007} &\textbf{37.45} &\textbf{0.0021} &\textbf{31.94} &\textbf{0.0076} \\
            \bottomrule
    \end{tabu}\vspace{-0.in}
     \caption{Comparison with super resolution methods with different upsampling scales on the BCR dataset~\cite{hou2021fast}. }
    \label{table:comp_sr}\vspace{-0.1in}
\end{table}

\noindent\textbf{Residual Densely-connected Swin Transformer block (RDST).} As shown in Figure~\ref{fig:network}, we feed the fused feature $X$ from XM to a sequence of $B=5$ residual densely-connected Swin Transformer blocks (RDST),
\begin{equation}
    F_{i-1}^b = f_{RDST}(F_{i-1}^{b-1}),
\end{equation}
where $f_{RDST}$ indicates the RDST block. We also use a short skip connection to combine the shallow feature $X_{i-1}$ with the deep feature $F_{i-1}^B$
\begin{equation}
    F_{i} = F_{i-1}^B + X_{i - 1}.
\end{equation}

We design RDST by combining the ideas of the Residual Densely-connected Network (RDN)~\cite{zhang2018residual} and Swin Transformer~\cite{liu2021swin}. 
We are specifically inspired by SwinIR~\cite{liang2021swinir} that explores Swin Transformers for image restoration tasks. It replaces traditional convolutional layers with Swin layers in residual blocks, allowing for the learning of more descriptive features and delivering impressive results. Taking inspiration from RDN~\cite{zhang2018residual}, we introduce RDST, where the convolution layers in densely-connected blocks are replaced with Swin layers.  As shown in Figure~\ref{fig:rdst}, RDST consists of a sequence of densely-connected Swin Transformer blocks and a local feature fusion block. For the densely-connected Swin Transformer blocks, we shift the windows. We also use a local skip connection to fuse the features from the shallow layer. 



\old{We predict the dense feature from the XDG by combining it with the shallow feature $F_0$ from the low-resolution rendering branch using a long skip connection,
\begin{equation}
    F_{DF} = F_N + F_0.
\end{equation}
}

\noindent\textbf{Upscale.} We adopt the pixel shuffle layer~\cite{shi2016real} to upscale the dense feature $F_{DF}$ to a high resolution feature. We also uses a $3\times3$ convolutional layer with 3 channels to predict the final high resolution image $I_{SR}$.
\begin{equation}
    I_{SR} = f_{conv}(f_{UP}(F_{DF})),
\end{equation}
where $f_{UP}$ indicates the operation of the pixel shuffle layer.

\noindent\textbf{Training details.} We adopt the robust loss to handle the prediction with a high dynamic range image~\cite{hou2021fast}.
\begin{equation}
\footnotesize
   \ell_r = \frac{1}{M}\sum_{p \in I_{HR}}\frac{|I_{HR}^p - I_{SR}^p|}{\beta + |I_{HR}^p - I_{SR}^p|},
\end{equation}
where $I_{HR}$ indicates the ground truth image. $M$ indicates the number of pixels. $\beta$ indicates the robust factor, which is set to 0.1.

We implement our network in PyTorch. We train our super resolution network on examples of size $256 \times 256$. We select Adam~\cite{kingma2014adam} with $\beta_1 = 0.9$, $\beta_2 = 0.999$ as the optimizer. The learning rate is set to 0.0001. We train the network for 400 epochs with a mini-batch size of 16 for our $4\times$ super resolution models, and we fine-tune our other models using the $4\times$ pretrained weights. It takes about one week to train a single model using 4 Nvidia A40 GPUs. We adopt the BCR dataset~\cite{hou2021fast} as the training dataset. BCR dataset contains 2449 images from 1463 scenes rendered by Blender Cycles. Following MSSPL~\cite{hou2021fast}, we use 2126 images from 1283 scenes for training, 193 images from 76 scenes for validation, and 130 images from 104 scenes for testing. 

\begin{figure}[t]
\footnotesize
  \includegraphics[width=0.48\textwidth]{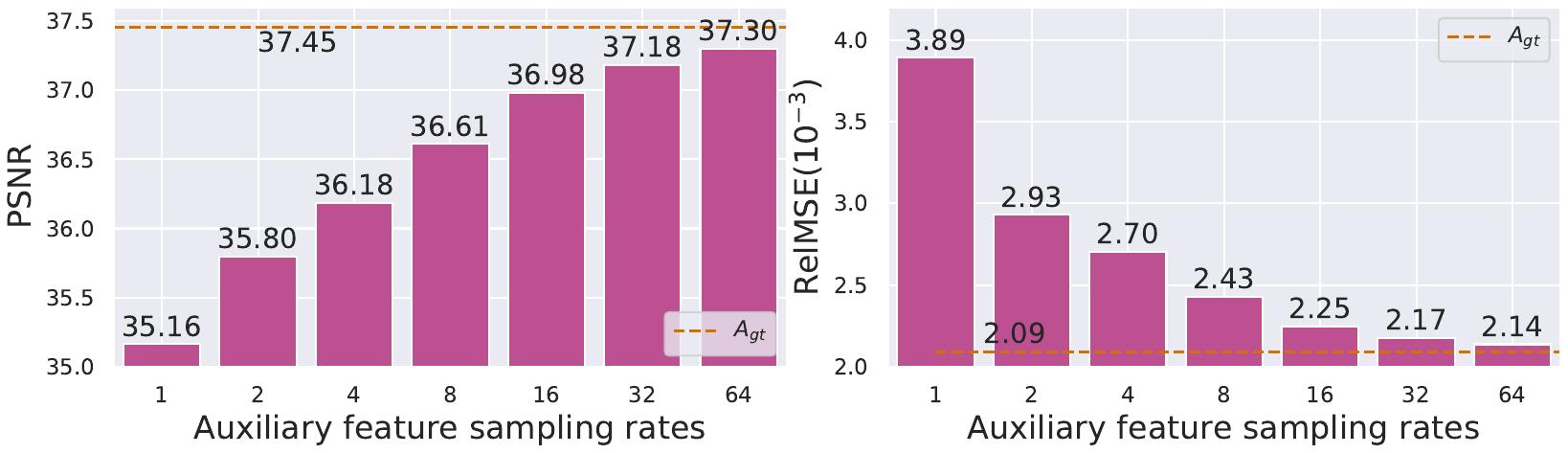}\vspace{-0.in}
  \caption{The effects of the sample rates used to generate fast-to-compute auxiliary features on the performance of our method. $A_{gt}$ indicates the ground truth  auxiliary features (4000spp).}
  \label{fig:aba_spp}\vspace{-0.2in}
\end{figure}

\vspace{0.05in}

\section{Experiments}
\label{sec:exp}
\vspace{0.05in}

We evaluate our network by quantitatively and qualitatively comparing them with state-of-the-art image super resolution methods and the Monte Carlo denoising methods on the BCR dataset~\cite{hou2021fast} and the Gharbi dataset~\cite{gharbi2019sample}. We also conduct the ablation study to examine our method. Following~\cite{hou2021fast}, we adopt Relative Mean Square Error (RelMSE) and PSNR to evaluate our methods in the scene linear color space and the sRGB space, respectively. Please refer to the supplementary material for an interactive demo that provides more results. 


\subsection{Comparison with Super Resolution Methods}
\label{sec::comp_sr}

            
             

We compare our method with state-of-the-art super-resolution methods, including EDSR~\cite{lim2017enhanced}, RCAN~\cite{zhang2018image}, and SwinIR~\cite{liang2021swinir}, a recent transformer-based approach, as well as the multiple sampling-based super resolution method MSSPL~\cite{hou2021fast}. We obtained the results of compared methods either from the authors~\cite{hou2021fast} or from finetuning the official models~\cite{lim2017enhanced,zhang2018image,liang2021swinir} on the BCR dataset.

As shown in Table~\ref{table:comp_sr}, our method outperforms super-resolution methods. This improvement can be largely attributed to the use of high-resolution auxiliary features to capture high-frequency visual details. For this experiment, we use the ground truth auxiliary features as they are fast to acquire. We also vary the number of samples used to generate these features in order to examine their effect on our method. As shown in Figure~\ref{fig:aba_spp}, while having more samples to generate these auxiliary features benefits our method, the features generated with only one sample per pixel allow our method to outperform the standard super-resolution methods. 



MSSPL takes both the low-resolution rendering and the high resolution noisy rendering as well as a wide variety of high resolution auxiliary features as input~\cite{hou2021fast}. In this test, the high resolution rendering and features are rendered with one sample per pixel. As shown in Figure~\ref{fig:aba_spp} and Table~\ref{table:comp_sr}, our method, when only using albedo and normal as auxiliary features obtained with one sample per pixel, can achieve 35.16 dB which is higher than MSSPL (34.27 dB) for the $\times4$ task, even though our method takes much less input information from the high resolution input.



\noindent \textbf{Speed and memory}. Table~\ref{table:speed} reports the speeds and the peak memory of the above methods. As our method is based on the Transformer, our method is slower than CNN-based methods, including EDSR~\cite{lim2017enhanced}, RCAN~\cite{zhang2018image} and MSSPL~\cite{hou2021fast}. This is consistent with many other works that Transformer tends to be slower than CNN~\cite{dosovitskiy2020image,pmlr-v139-touvron21a,liang2021swinir}. Compared to Transformer-based method SwinIR, our method is slightly faster.
We also compare the peak memory to produce a $1024\times1024$ image in Table~\ref{table:speed}. Our method uses less peak memory than EDSR and MSSPL but more memory than RCAN and SwinIR.

\begin{figure*}[thp]
    \newlength\indentspace
    \setlength{\indentspace}{-3.8mm}
   
        \begin{tabular}{cc}

        \hspace{-3.8mm}
        \begin{adjustbox}{valign=t}
        \tiny
            \begin{tabular}{c}
              \includegraphics[width=0.335\textwidth]{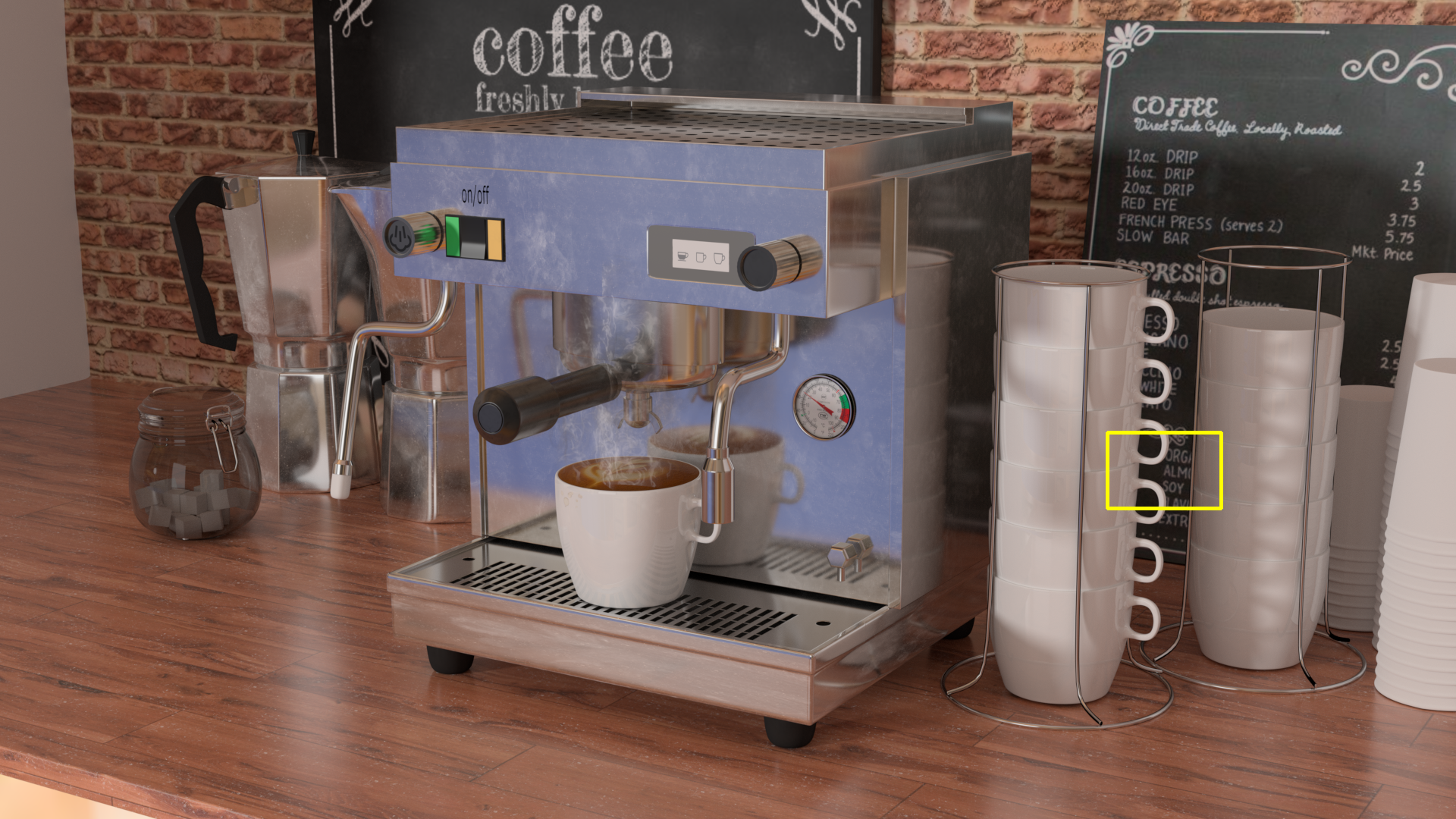}
                \\
                Ground Truth
            \end{tabular}
        \end{adjustbox}
        \hspace{-7.8mm}
        &
        \begin{adjustbox}{valign=t}
        \tiny
            \begin{tabular}{ccccc}
                \includegraphics[width=\subsmallfigsize\textwidth]{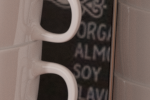} \hspace{\indentspace} &
                \includegraphics[width=\subsmallfigsize\textwidth]{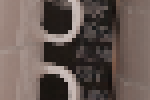} \hspace{\indentspace} &
                \includegraphics[width=\subsmallfigsize\textwidth]{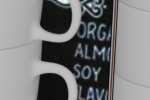} \hspace{\indentspace} &
                \includegraphics[width=\subsmallfigsize\textwidth]{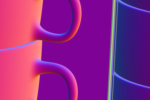} \hspace{\indentspace} &
                \includegraphics[width=\subsmallfigsize\textwidth]{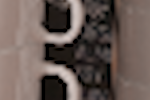}
                \\
                GT(PSNR$\uparrow$/RelMSE$\downarrow$) \hspace{\indentspace} &
                LR\hspace{\indentspace} &
                Albedo \hspace{\indentspace} &
                Normal \hspace{\indentspace} &
                Bicubic(25.73/0.0230)
                \\
                \includegraphics[width=\subsmallfigsize\textwidth]{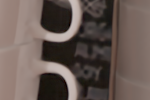} \hspace{\indentspace} &
                \includegraphics[width=\subsmallfigsize\textwidth]{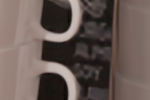} \hspace{\indentspace} &
                \includegraphics[width=\subsmallfigsize\textwidth]{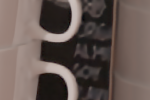} \hspace{\indentspace} &
                \includegraphics[width=\subsmallfigsize\textwidth]{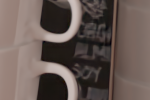} \hspace{\indentspace} &
                \includegraphics[width=\subsmallfigsize\textwidth]{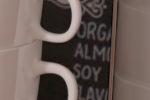}
                \\ 
                EDSR(31.84/0.0048) \hspace{\indentspace} &
                RCAN(31.77/0.0049) \hspace{\indentspace} &
                SwinIR(31.23/0.0056) \hspace{\indentspace} &
                MSSPL(35.70/0.0018) \hspace{\indentspace} &
                Ours(\textbf{38.82}/\textbf{0.0009})
                \\
            \end{tabular}
        \end{adjustbox} 
        \\

        \hspace{-3.8mm}
        \begin{adjustbox}{valign=t}
        \tiny
            \begin{tabular}{c}
              \includegraphics[width=0.335\textwidth]{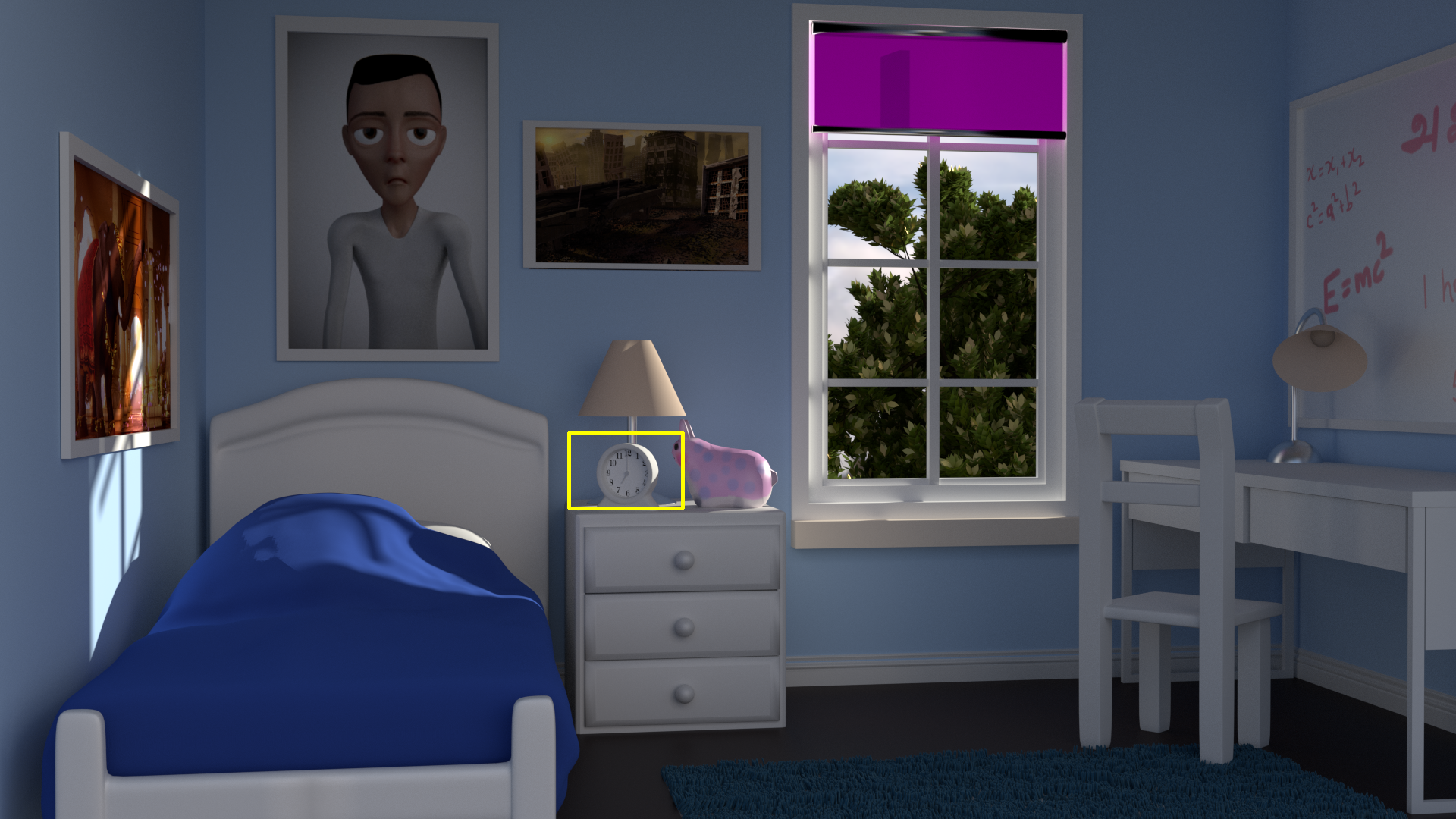}
                \\
                Ground Truth
            \end{tabular}
        \end{adjustbox}
        \hspace{-7.8mm}
        &
        \begin{adjustbox}{valign=t}
        \tiny
            \begin{tabular}{ccccc}
                \includegraphics[width=\subsmallfigsize\textwidth]{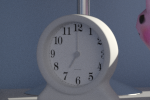} \hspace{\indentspace} &
                \includegraphics[width=\subsmallfigsize\textwidth]{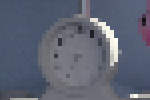} \hspace{\indentspace} &
                \includegraphics[width=\subsmallfigsize\textwidth]{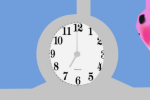} \hspace{\indentspace} &
                \includegraphics[width=\subsmallfigsize\textwidth]{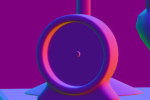} \hspace{\indentspace} &
                \includegraphics[width=\subsmallfigsize\textwidth]{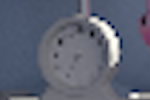}
                \\
                GT(PSNR$\uparrow$/RelMSE$\downarrow$) \hspace{\indentspace} &
                LR\hspace{\indentspace} &
                Albedo \hspace{\indentspace} &
                Normal \hspace{\indentspace} &
                Bicubic(27.59/0.0171)
                \\
                \includegraphics[width=\subsmallfigsize\textwidth]{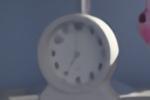} \hspace{\indentspace} &
                \includegraphics[width=\subsmallfigsize\textwidth]{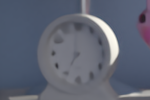} \hspace{\indentspace} &
                \includegraphics[width=\subsmallfigsize\textwidth]{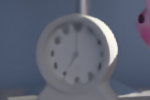} \hspace{\indentspace} &
                \includegraphics[width=\subsmallfigsize\textwidth]{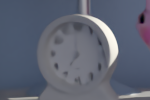} \hspace{\indentspace} &
                \includegraphics[width=\subsmallfigsize\textwidth]{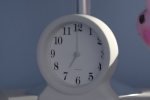}
                \\ 
                EDSR(33.69/0.0030) \hspace{\indentspace} &
                RCAN(33,53/0.0033) \hspace{\indentspace} &
                SwinIR(33.66/0.0029) \hspace{\indentspace} &
                MSSPL(38.12/0.0008) \hspace{\indentspace} &
                Ours(\textbf{39.98}/\textbf{0.0006})
                \\
            \end{tabular}
        \end{adjustbox} 
        \\

        \hspace{-3.8mm}
        \begin{adjustbox}{valign=t}
        \tiny
            \begin{tabular}{c}
              \includegraphics[width=0.335\textwidth]{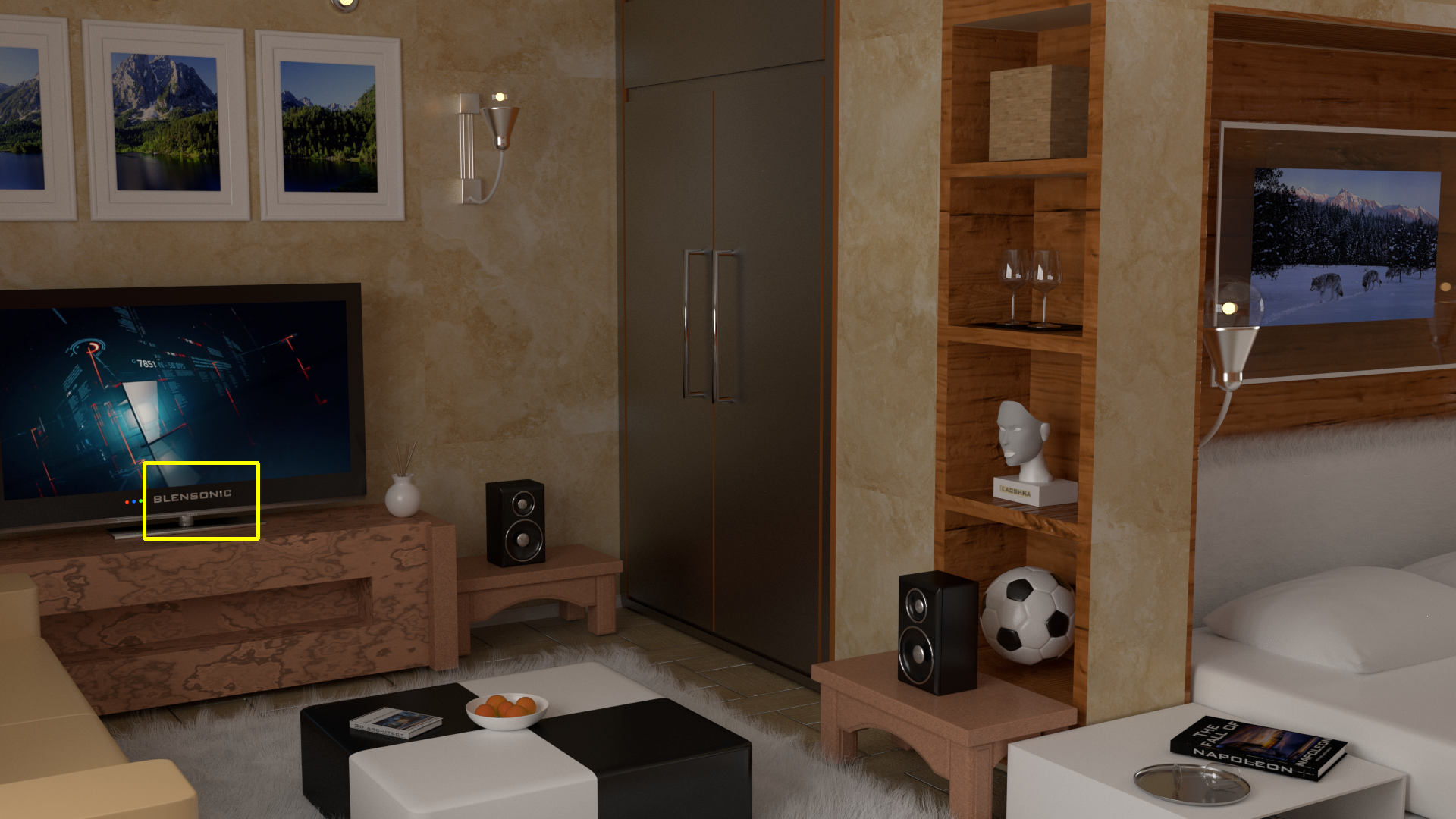}
                \\
                Ground Truth
            \end{tabular}
        \end{adjustbox}
        \hspace{-7.8mm}
        &
        \begin{adjustbox}{valign=t}
        \tiny
            \begin{tabular}{ccccc}
                \includegraphics[width=\subsmallfigsize\textwidth]{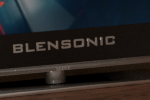} \hspace{\indentspace} &
                \includegraphics[width=\subsmallfigsize\textwidth]{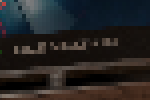} \hspace{\indentspace} &
                \includegraphics[width=\subsmallfigsize\textwidth]{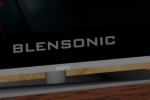} \hspace{\indentspace} &
                \includegraphics[width=\subsmallfigsize\textwidth]{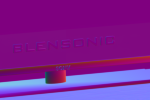} \hspace{\indentspace} &
                \includegraphics[width=\subsmallfigsize\textwidth]{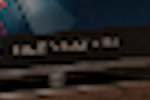}
                \\
                GT(PSNR$\uparrow$/RelMSE$\downarrow$) \hspace{\indentspace} &
                LR\hspace{\indentspace} &
                Albedo \hspace{\indentspace} &
                Normal \hspace{\indentspace} &
                Bicubic(28.45/0.0116)
                \\
                \includegraphics[width=\subsmallfigsize\textwidth]{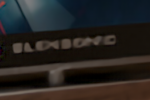} \hspace{\indentspace} &
                \includegraphics[width=\subsmallfigsize\textwidth]{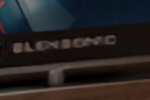} \hspace{\indentspace} &
                \includegraphics[width=\subsmallfigsize\textwidth]{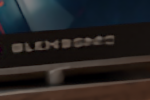} \hspace{\indentspace} &
                \includegraphics[width=\subsmallfigsize\textwidth]{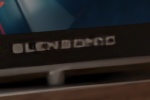} \hspace{\indentspace} &
                \includegraphics[width=\subsmallfigsize\textwidth]{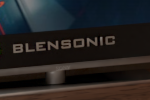}
                \\ 
                EDSR(34.34/0.0020) \hspace{\indentspace} &
                RCAN(34.32/0.0021) \hspace{\indentspace} &
                SwinIR(34.20/0.0021) \hspace{\indentspace} &
                MSSPL(37.21/0.0010) \hspace{\indentspace} &
                Ours(\textbf{41.28}/\textbf{0.0004})
                \\
            \end{tabular}
        \end{adjustbox} 
        \\
        
    \end{tabular}\vspace{-0.in}
    \caption{
        Visual comparison with super-resolution methods on the BCR dataset~\cite{hou2021fast}. 
    }
    \label{fig:comp_sr}
\end{figure*}

\begin{figure*}[thp]

    \setlength{\indentspace}{-3.8mm}
        \begin{tabular}{cc}
        \hspace{-4mm}
        \begin{adjustbox}{valign=t}
        \tiny
            \begin{tabular}{c}
              \includegraphics[width=0.385\textwidth]{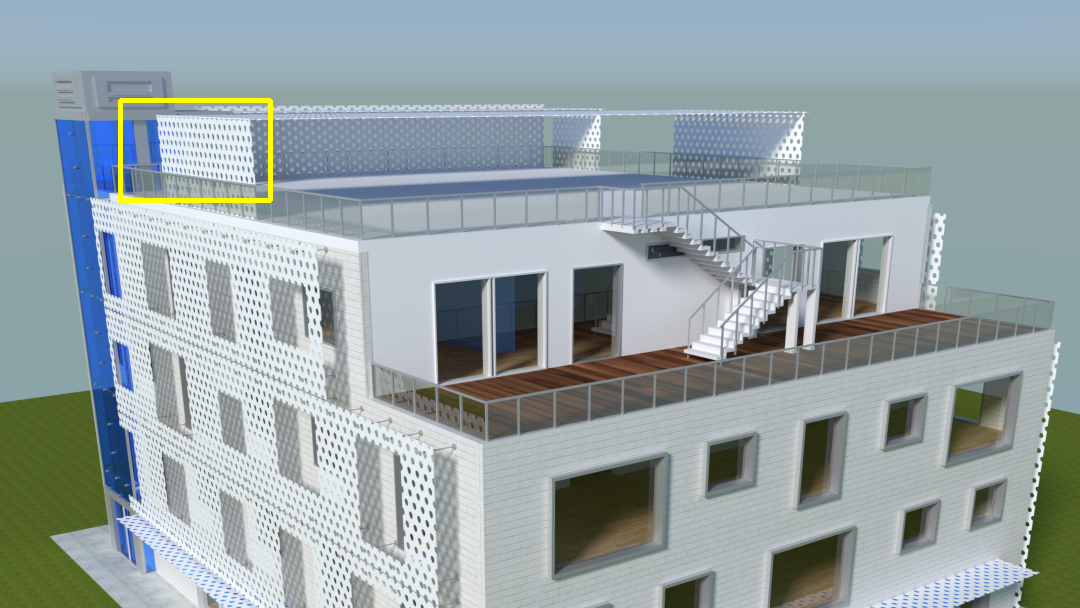}
                \\
                Ground Truth
            \end{tabular}
        \end{adjustbox}
        \hspace{-8.1mm}
        &
        \begin{adjustbox}{valign=t}
        \tiny
            \begin{tabular}{cccc}
                \includegraphics[width=\subfigfoursize\textwidth]{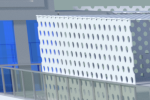} \hspace{\indentspace} &
                \includegraphics[width=\subfigfoursize\textwidth]{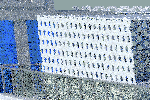} \hspace{\indentspace} &
                \includegraphics[width=\subfigfoursize\textwidth]{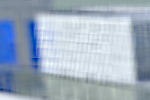} \hspace{\indentspace} &
                \includegraphics[width=\subfigfoursize\textwidth]{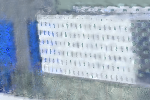}
                \\
                GT (PSNR$\uparrow$/RelMSE$\downarrow$) \hspace{\indentspace} &
                2spp \hspace{\indentspace} &
                Bitterli(25.53/0.0290) \hspace{\indentspace} &
                KPCN(25.38/0.0611)
                \\
                \includegraphics[width=\subfigfoursize\textwidth]{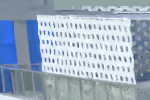} \hspace{\indentspace} &
                \includegraphics[width=\subfigfoursize\textwidth]{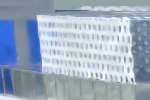} \hspace{\indentspace} &
                \includegraphics[width=\subfigfoursize\textwidth]{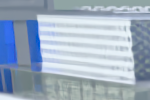} \hspace{\indentspace} &
                \includegraphics[width=\subfigfoursize\textwidth]{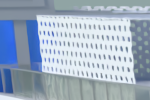}
                \\ 
                KPCN-ft(28.20/0.0416) \hspace{\indentspace} &
                Gharbi(28.27/0.0122) \hspace{\indentspace} &
                MSSPL(31.02/0.0382) \hspace{\indentspace} &
                Ours(\textbf{32.70}/\textbf{0.0035})
                \\
            \end{tabular}
        \end{adjustbox} 
        \\

         \hspace{-4mm}
         \begin{adjustbox}{valign=t}
        \tiny
            \begin{tabular}{c}
              \includegraphics[width=0.385\textwidth]{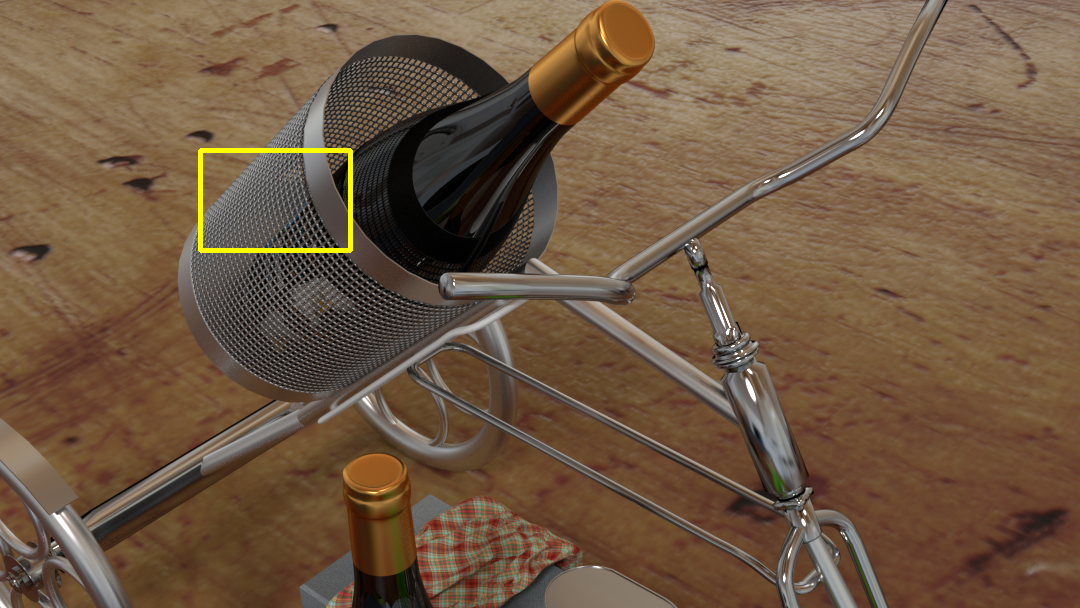}
                \\
                Ground Truth
            \end{tabular}
        \end{adjustbox}
        \hspace{-8.1mm}
        &
        \begin{adjustbox}{valign=t}
        \tiny
            \begin{tabular}{cccc}
                \includegraphics[width=\subfigfoursize\textwidth]{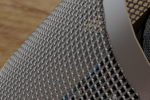} \hspace{\indentspace} &
                \includegraphics[width=\subfigfoursize\textwidth]{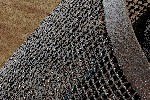} \hspace{\indentspace} &
                \includegraphics[width=\subfigfoursize\textwidth]{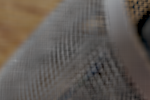} \hspace{\indentspace} &
                \includegraphics[width=\subfigfoursize\textwidth]{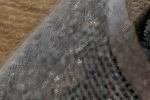}
                \\
                GT (PSNR$\uparrow$/RelMSE$\downarrow$) \hspace{\indentspace} &
                2spp \hspace{\indentspace} &
                Bitterli(26.82/0.0131) \hspace{\indentspace} &
                KPCN(24.50/0.0984)
                \\
                \includegraphics[width=\subfigfoursize\textwidth]{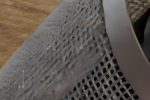} \hspace{\indentspace} &
                \includegraphics[width=\subfigfoursize\textwidth]{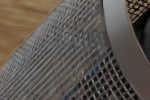} \hspace{\indentspace} &
                \includegraphics[width=\subfigfoursize\textwidth]{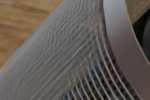} \hspace{\indentspace} &
                \includegraphics[width=\subfigfoursize\textwidth]{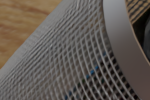}
                \\ 
                KPCN-ft(30.66/0.0461) \hspace{\indentspace} &
                Gharbi(30.07/0.0050) \hspace{\indentspace} &
                MSSPL(31.77/0.0519) \hspace{\indentspace} &
                Ours(\textbf{32.86}/\textbf{0.0029})
                \\
            \end{tabular}
        \end{adjustbox} 
        \\
    \end{tabular}\vspace{-0.in}
    \caption{
        Visual comparison with denoising methods on the BCR dataset~\cite{hou2021fast}. 
    }
    \label{fig:comp_bcr}

\end{figure*}

\begin{figure*}[thp]
    \setlength{\indentspace}{-3.8mm}
   
        \begin{tabular}{cc}
        \hspace{-4mm}
        \begin{adjustbox}{valign=t}
        \tiny
            \begin{tabular}{c}
              \includegraphics[width=0.323\textwidth]{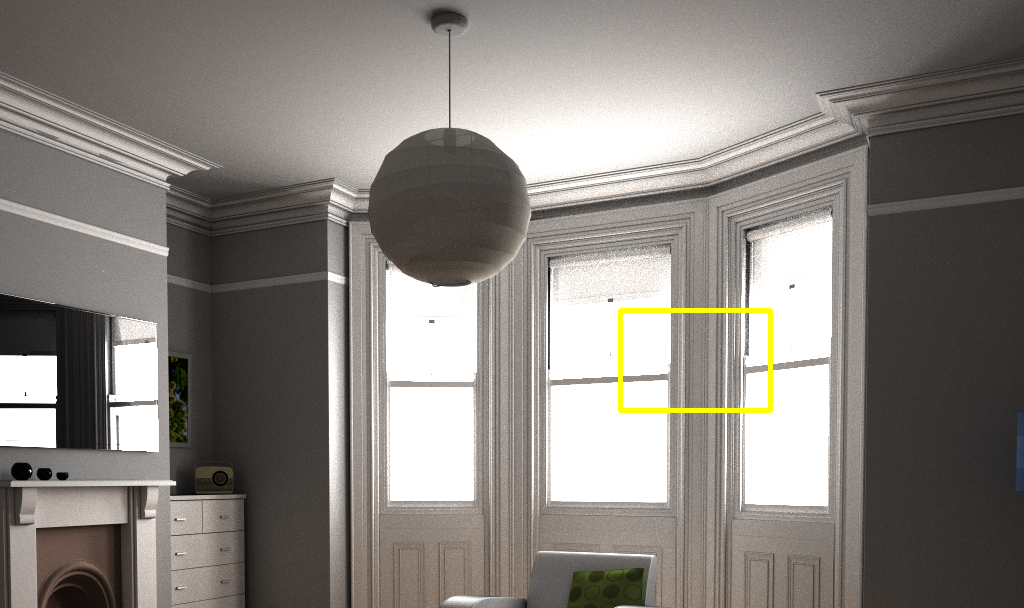}
                \\
                Ground Truth
            \end{tabular}
        \end{adjustbox}
        \hspace{-8mm}
        &
        \begin{adjustbox}{valign=t}
        \tiny
            \begin{tabular}{ccccc}
                \includegraphics[width=\subfigsbmcsize\textwidth]{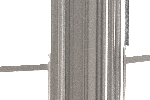} \hspace{\indentspace} &
                \includegraphics[width=\subfigsbmcsize\textwidth]{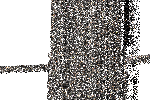} \hspace{\indentspace} &
                \includegraphics[width=\subfigsbmcsize\textwidth]{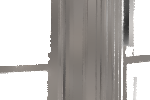} \hspace{\indentspace} &
                \includegraphics[width=\subfigsbmcsize\textwidth]{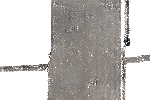} \hspace{\indentspace} &
                \includegraphics[width=\subfigsbmcsize\textwidth]{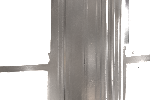} 
                \\
                GT (PSNR$\uparrow$/RelMSE$\downarrow$) \hspace{\indentspace} &
                4spp \hspace{\indentspace} &
                Sen(22.12/0.0410) \hspace{\indentspace} &
                Rousselle(23.87/0.0915) \hspace{\indentspace} &
                Kalantari(25.02/0.0444) 
                \\
                \includegraphics[width=\subfigsbmcsize\textwidth]{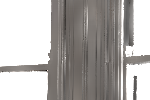} \hspace{\indentspace} &
                \includegraphics[width=\subfigsbmcsize\textwidth]{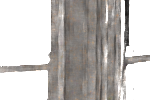} \hspace{\indentspace} &
                \includegraphics[width=\subfigsbmcsize\textwidth]{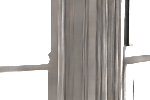} \hspace{\indentspace} &
                \includegraphics[width=\subfigsbmcsize\textwidth]{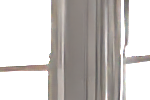} \hspace{\indentspace} &
                \includegraphics[width=\subfigsbmcsize\textwidth]{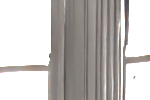}
                \\ 
                Bitterli(21.76/0.0546) \hspace{\indentspace} &
                KPCN-ft(25.78/0.0175) \hspace{\indentspace} &
                Gharbi(27.04/\textbf{0.0105}) \hspace{\indentspace} &
                MSSPL(27.89/0.0166) \hspace{\indentspace} &
                Ours $\times2$(\textbf{27.90}/0.0141)
                \\
            \end{tabular}
        \end{adjustbox} 
        \\

        \hspace{-4mm}
        \begin{adjustbox}{valign=t}
        \tiny
            \begin{tabular}{c}
              \includegraphics[width=0.323\textwidth]{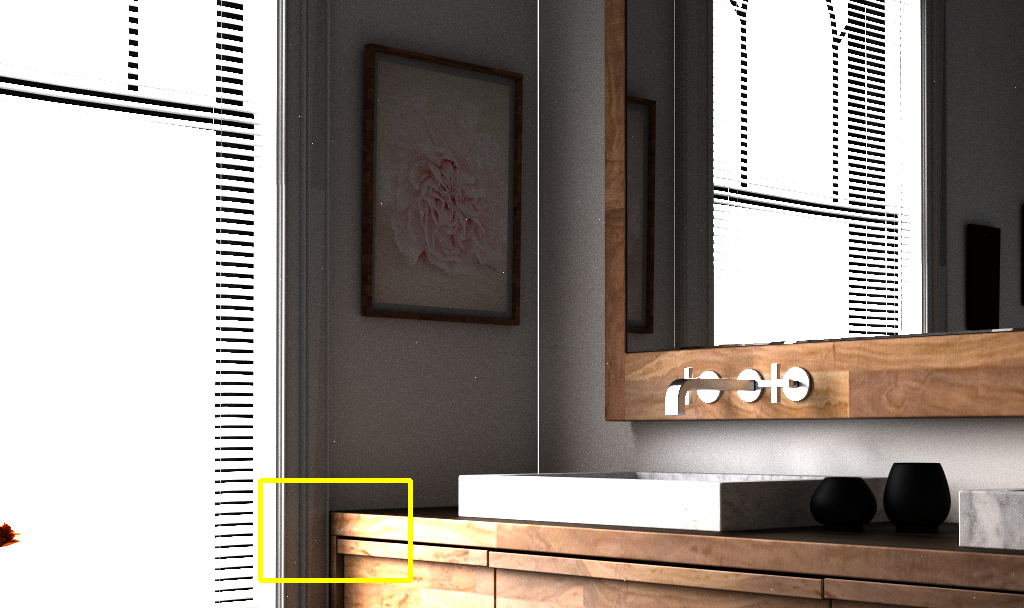}
                \\
                Ground Truth
            \end{tabular}
        \end{adjustbox}
        \hspace{-8mm}
        &
        \begin{adjustbox}{valign=t}
        \tiny
            \begin{tabular}{ccccc}
                \includegraphics[width=\subfigsbmcsize\textwidth]{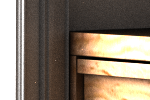} \hspace{\indentspace} &
                \includegraphics[width=\subfigsbmcsize\textwidth]{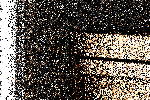} \hspace{\indentspace} &
                \includegraphics[width=\subfigsbmcsize\textwidth]{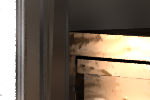} \hspace{\indentspace} &
                \includegraphics[width=\subfigsbmcsize\textwidth]{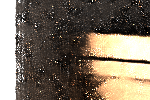} \hspace{\indentspace} &
                \includegraphics[width=\subfigsbmcsize\textwidth]{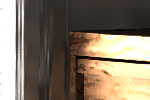} 
                \\
                GT (PSNR$\uparrow$/RelMSE$\downarrow$) \hspace{\indentspace} &
                4spp \hspace{\indentspace} &
                Sen(20.31/46.534) \hspace{\indentspace} &
                Rousselle(18.74/1.001) \hspace{\indentspace} &
                Kalantari(20.72/9.876) 
                \\
                \includegraphics[width=\subfigsbmcsize\textwidth]{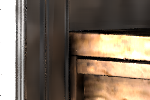} \hspace{\indentspace} &
                \includegraphics[width=\subfigsbmcsize\textwidth]{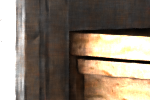} \hspace{\indentspace} &
                \includegraphics[width=\subfigsbmcsize\textwidth]{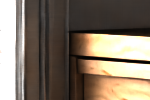} \hspace{\indentspace} &
                \includegraphics[width=\subfigsbmcsize\textwidth]{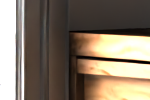} \hspace{\indentspace} &
                \includegraphics[width=\subfigsbmcsize\textwidth]{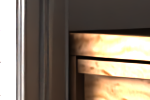}
                \\ 
                Bitterli(21.74/16.060) \hspace{\indentspace} &
                KPCN-ft(23.22/0.105) \hspace{\indentspace} &
                Gharbi(26.01/\textbf{0.0418}) \hspace{\indentspace} &
                MSSPL(24.74/2.689) \hspace{\indentspace} &
                Ours $\times2$(\textbf{26.18}/1.112)
                \\
            \end{tabular}
        \end{adjustbox} 
        \\
        
    \end{tabular}\vspace{-0.in}
    \caption{
        Visual comparison with denoising methods on the Gharbi dataset~\cite{gharbi2019sample}. 
    }
   \vspace{-0.in}
    \label{fig:comp_sbmc}
\end{figure*}

\subsection{Comparison with Denoising Methods}

\begin{table}[t]
\setlength{\tabcolsep}{3.2pt}
    \centering
    \small
    \begin{tabular}{lcccccc}
            \toprule
             &Scale & EDSR &RCAN &SwinIR  &MSSPL &Ours \\ \midrule
            Runtime(ms) &$\times4$  &503.96     &280.51   &1149.25 &125.24  &1009.08\\
            Peak memory(MB) &$\times4$ &2493.9    &672.1  &806.0  &739.70 &941.3\\
            Peak memory(MB) &$\times8$   & 2375.6   &621.3 &659.0 & 1010.1 &803.8 \\
            Peak memory(MB) &$\times16$  & 2359.7 & 615.4 & 608.4 & 1008.0 & 783.4 \\
            \bottomrule
    \end{tabular}\vspace{-0.in}
     \caption{Comparison of runtime cost and peak memory with super resolution methods to produce a $1024\times1024$ image on an Nvidia Titan XP.} \vspace{-0.1in}
     \label{table:speed}
\end{table}

\begin{table}
    \centering
    \small
    \begin{tabular}{P{1.5cm}P{0.45cm}P{0.9cm}P{0.45cm}P{0.9cm}P{0.45cm}P{0.9cm}}
            \toprule
            \multirow{2}[2]{*}{Method} & \multicolumn{2}{c}{2spp} & \multicolumn{2}{c}{4spp} & \multicolumn{2}{c}{8spp}  
            \\ \cmidrule(lr){2-3}  \cmidrule(lr){4-5}  \cmidrule(lr){6-7} 
             &PSNR &RelMSE &PSNR &RelMSE &PSNR &RelMSE 
             \\ \midrule
             Input &18.12 &0.2953 &21.51 &0.1400 &24.75 &0.0646 \\
             KPCN &25.87 &0.0390 &27.31 &0.0299 &28.11 &0.0276 \\
             KPCN-ft &31.03 &0.0078 &33.69 &0.0043 &35.83 &0.0026 \\
             Bitterli &26.67 &0.0293 &27.22 &0.0252 &27.45 &0.0226 \\
             Gharbi &30.73 & 0.0068 &31.61 &0.0057 &32.29 &0.0050\\
             MSSPL$\times2$ &33.27 &0.0044 &35.15 &0.0027 &36.74 &0.0019\\
             MSSPL$\times4$ &33.94 &0.0039 &35.21 &0.0028 &36.31 &0.0022\\
             MSSPL$\times8$ &31.37 &0.0075 &32.35 &0.0057 &33.14 &0.0049\\
             AdvMC-ft &30.33   &-  &32.30  &- &33.69 &- \\
             MCSA-ft &32.68  &0.0049 &34.81 &0.0031 &36.61 &0.0021 \\
              \midrule
            \multirow{2}[0]{*}{Ours$\times1$}  & \multicolumn{2}{c}{(1 - 1)}  & \multicolumn{2}{c}{( 2 -  2)}  & \multicolumn{2}{c}{(4 - 4)}  \\
             &31.04 &0.0078 &34.67 &0.0030 &36.62 &0.0020 \\
             \multirow{2}[0]{*}{Ours$\times2$}  & \multicolumn{2}{c}{(4 - 1)}  & \multicolumn{2}{c}{( 8 -  2)}  & \multicolumn{2}{c}{(16 - 4)}  \\
             &\textbf{34.12} &\textbf{0.0035} &\textbf{35.49} &\textbf{0.0026} &\textbf{37.09} &\textbf{0.0018} \\
             
             \multirow{2}[0]{*}{Ours$\times4$}   & \multicolumn{2}{c}{(16 - 1)}  & \multicolumn{2}{c}{(32 - 2)}  & \multicolumn{2}{c}{(64 - 4)} \\ 
             &34.08 &0.0046 &35.06 &0.0034 &35.77 &0.0029 \\
             \multirow{2}[0]{*}{Ours$\times8$}   & \multicolumn{2}{c}{(64 - 1)}  & \multicolumn{2}{c}{(128 - 2)}  & \multicolumn{2}{c}{(250 - 4)} \\ 
             &31.10 &0.0095 &31.36 &0.0093 &31.62 &0.0093\\
            \bottomrule
    \end{tabular}\vspace{-0.in}
        \caption{Comparison on the BCR dataset~\cite{hou2021fast}.  (4 - 1) indicates that our method takes a 4-spp  low resolution rendering and 1 spp fast-to-compute auxiliary feature as input. MSSPL also takes the 1-spp high-resolution rendering as input, including the diffusion and specular layers. Our method takes less shading information compared to MSSPL~\cite{hou2021fast}.} \label{table:comp_bcr} \vspace{-0.in}
        
\end{table}


\begin{table}
    \centering
    \small
    \begin{tabular}{P{1.5cm}P{0.45cm}P{0.9cm}P{0.45cm}P{0.9cm}P{0.45cm}P{0.9cm}}
            \toprule
            \multirow{2}[2]{*}{Method} & \multicolumn{2}{c}{4 spp} & \multicolumn{2}{c}{8 spp}  &\multicolumn{2}{c}{16 spp} 
            \\ \cmidrule(lr){2-3}  \cmidrule(lr){4-5}  \cmidrule(lr){6-7} 
             &PSNR &RelMSE &PSNR &RelMSE &PSNR &RelMSE 
             \\ \midrule
             Input &19.58 &17.5358 &21.91 &7.5682 &24.17 &11.2189 \\
             Sen &28.23 &1.0484 &28.00 &0.5744 &27.64 &0.3396\\
             Rousselle &30.01 &1.9407 &32.32 &1.9660 &34.36 &1.9446 \\
             Kalantari &31.33 &1.5573 &33.00 &1.6635 &34.43 &1.8021 \\
             Bitterli & 28.98 &1.1024 &30.92 &0.9297 &32.40 &0.9640 \\
             KPCN &29.75 &1.0616 &30.56 &7.0774 &31.00 &20.2309 \\
             KPCN-ft &29.86 &0.5004 &31.66 &0.8616 &33.39 &0.2981 \\ 
             Gharbi &33.11 &\textbf{0.0486} &34.45 &\textbf{0.0385} &35.36 &\textbf{0.0318}\\ 
             MSSPL$\times2$ &34.02 &1.5025 &35.30 &1.4902 &\textbf{36.43} &1.4748 \\
             MSSPL$\times4$ &33.94 &5.5586 &35.22 &5.6781 &35.97 &5.7436 \\
             MSSPL$\times8$ &31.56 &3.7228 &32.60 &4.2300 &33.22 &4.5045 \\
             \midrule
             \multirow{2}[0]{*}{Ours$\times1$} & \multicolumn{2}{c}{(2 - 2)}  & \multicolumn{2}{c}{(4 - 4)}  & \multicolumn{2}{c}{(8 - 8)}  \\ 
             &27.41 &0.3438 &30.39 &0.3092 &32.88 &0.3062 \\
             \multirow{2}[0]{*}{Ours$\times2$} & \multicolumn{2}{c}{(8 - 2)}  & \multicolumn{2}{c}{(16 - 4)}  & \multicolumn{2}{c}{(32 - 8)}  \\ 
             &\textbf{34.29} &2.2587 &\textbf{35.47} &1.5480 &36.37 &1.5417 \\
             \multirow{2}[0]{*}{Ours$\times4$} & \multicolumn{2}{c}{(32 - 2)}  & \multicolumn{2}{c}{(64 - 4)}  & \multicolumn{2}{c}{(128 - 8)}  \\
             &34.26 &20.7861 &35.12 &29.0364 &35.52 & 28.1264\\
             \multirow{2}[0]{*}{Ours$\times8$} & \multicolumn{2}{c}{(128 - 2)}  & \multicolumn{2}{c}{(16 -  8)}  & \multicolumn{2}{c}{(32 - 16)}  \\
             &31.57 &1.3474 &31.26 &1.1718 &31.51 &1.0940\\
            \bottomrule
    \end{tabular}\vspace{-0.in}
     \caption{Comparison on the Gharbi dataset~\cite{gharbi2019sample}. We directly test our models pretrained on the BCR dataset without finetuning.} \vspace{-0.in}
     \label{table:comp_sbmc}
\end{table}


\begin{table}
    \centering
    \small
    \begin{tabular}{P{0.8cm}P{0.8cm}P{0.45cm}P{0.8cm}P{0.4cm}P{0.8cm}P{0.4cm}P{0.8cm}}
            \toprule
            \multirow{2}[2]{*}{Method} & \multirow{2}[2]{*}{Buffer} & \multicolumn{2}{c}{2spp} & \multicolumn{2}{c}{4spp} & \multicolumn{2}{c}{8spp}  
            \\ \cmidrule(lr){3-4}  \cmidrule(lr){5-6}  \cmidrule(lr){7-8} 
             & &PSNR &RelMSE &PSNR &RelMSE &PSNR &RelMSE 
             \\ \midrule
             \multirow{2}[0]{*}{MSSPL}  & \multirow{2}[0]{*}{Full} & \multicolumn{2}{c}{(16 - 1)}  & \multicolumn{2}{c}{(32 - 2)}  & \multicolumn{2}{c}{(64 - 4)} \\ 
             & &33.94 &0.0039 &35.21 &0.0028 &36.31 &0.0022 \\ 

             \multirow{2}[0]{*}{MSSPL}  & \multirow{2}[0]{*}{Fast} & \multicolumn{2}{c}{(16 - 1)}  & \multicolumn{2}{c}{(32 - 2)}  & \multicolumn{2}{c}{(64 - 4)} \\ 
             & &32.32 &0.0079 &33.42 &0.0060 &33.96 &0.0050 \\ 

             \multirow{2}[0]{*}{Ours}  & \multirow{2}[0]{*}{Full} & \multicolumn{2}{c}{(16 - 1)}  & \multicolumn{2}{c}{(32 - 2)}  & \multicolumn{2}{c}{(64 - 4)} \\ 
             & &34.84 &0.0033 &36.01 &0.0024 &37.22 &0.0018 \\ 

             \multirow{2}[0]{*}{Ours}  & \multirow{2}[0]{*}{Fast} & \multicolumn{2}{c}{(16 - 1)}  & \multicolumn{2}{c}{(32 - 2)}  & \multicolumn{2}{c}{(64 - 4)} \\ 
             & &34.08 &0.0046 &35.06 &0.0034 &35.77 &0.0029 \\ 
            \bottomrule
    \end{tabular}\vspace{-0.in}
        \caption{Ablation study w.r.t. MSSPL~\cite{hou2021fast} on the BCR dataset~\cite{hou2021fast}. We compare their performance using fast-to-compute auxiliary features layers (``Fast'') and full auxiliary feature layers (``Full'').} \label{table:aba_fast_buffer} \vspace{-0.in}   
\end{table}

\begin{table}[thp]
    \centering
    \small
    \begin{tabular}{ccccc}
            \toprule
            Auxiliary Layer & None & Normal  &Albedo & Normal + Albedo \\ \midrule
            PSNR   &30.49     &34.85   &36.42   &\textbf{37.45}\\
            RelMSE &0.0141    &0.0042  &0.0030  &\textbf{0.0021}\\
            \bottomrule
    \end{tabular} 
     \caption{The effects of input fat-to-compute auxiliary feature layers on the BCR dataset~\cite{hou2021fast}.}  
     \label{table:aux_layers}

    \vspace{0.1in}
    \small
    \setlength{\tabcolsep}{17pt}
    \begin{tabular}{cccc}
            \toprule
            Method &AdvMC-ft &MCSA  &Ours \\ \midrule
            PSNR   &27.96     &30.01 &\textbf{34.12}\\
            LPIPS  &0.320    &0.202  &\textbf{0.090}\\
            \bottomrule
    \end{tabular}\vspace{-0.in}
     \caption{The effects of network architectures on the BCR dataset~\cite{hou2021fast}. AdvMC-ft~\cite{xu2019adversarial} and MCSA~\cite{yu2021monte} take 1spp RGB and 1spp auxiliary buffers as inputs.  Our method takes 4-spp low-resolution RGB ($\times2$, effectively the same sampling rate as 1-spp at the high resolution ) and 1-spp high-resolution auxiliary buffers)} \vspace{-0.1in}
     \label{table:arch_denoise}
\end{table}

We compare our methods to the state-of-the-art Monte Carlo denoising methods, including Sen~\cite{sen2012filtering}, Rousselle~\cite{rousselle2011adaptive}, Kalantari~\cite{kalantari2015machine}, Bitterli~\cite{bitterli2016nonlinearly}, KPCN\cite{bako2017kernel}, Gharbi~\cite{gharbi2019sample}, MSSPL~\cite{hou2021fast}, AdvMC~\cite{xu2019adversarial}, and MCSA~\cite{yu2021monte}. Table~\ref{table:comp_bcr} and Table~\ref{table:comp_sbmc} report results on the BCR dataset~\cite{hou2021fast} and the Gharbi dataset~\cite{gharbi2019sample}, respectively. We obtain the results of the comparing methods either from their authors~\cite{hou2021fast} or from their project websites~\cite{gharbi2019sample}.  MSSPL~\cite{hou2021fast} was trained on the BCR dataset. For KPCN~\cite{bako2017kernel}, AdvMC~\cite{xu2019adversarial}, and MCSA~\cite{yu2021monte}, we finetuned their official models on the BCR dataset using their official training scripts. For our model, We trained a distinct model for each scale and sampling count.

As most denoising methods do not take high-resolution auxiliary features as input, we follow MSSPL~\cite{hou2021fast} to compute the average spp for our method and MSSPL as $app_{avg} = spp_{LR}/s^2 + spp_{HR}$, where $s$ indicates the scale. $spp_{LR}$ and $spp_{HR}$ indicate the sampling rates for the low-resolution and high-resolution inputs, respectively. In our case, we take the sampling rates for the auxiliary features as $spp_{HR}$. We would like to note that this measurement of spp is \textbf{unfair} to our method, as our method 
only uses high-resolution albedo and normal features which takes much less time than rendering all the shading layers to obtain the high-resolution rendering as done in MSSPL.


As shown in Table~\ref{table:comp_bcr}, our method generates better results than the state-of-the-art methods on the BCR dataset~\cite{hou2021fast}. Ours $\times2$ model wins 0.18dB, 0.28dB, and 0.35 dB in terms of PSNR on 2spp, 4sppp, and 8spp, respectively.


We also conduct our experiments on $\times16$ scale. On the one hand, with $\times16$, our method produces worse results than MSSPL because MSSPL uses the high-resolution RGB image as input that is not available to our method. While the high-resolution RGB input to MSSPL is rendered at a low sampling rate, it still provides useful information. As shown in the existing literature on Monte Carlo denoising, even the rendering result at 1 spp can be denoised to a reasonable quality. At such a high upsampling rate of $\times16$, super-resolution is very difficult. On the other hand, in practice, given a target overall spp rate, our method can select an optimal combination of (spp rate, super-resolution scale) that outperforms MSSPL and other methods, as shown in Table~\ref{table:comp_bcr}. In practice, $\times16$ will not be used for rendering by either MSSPL or our method to achieve an overall target spp as it produces the worst results among alternative combinations of spp rate and super-resolution scale.

Figure \ref{fig:comp_bcr} shows the visual comparisons. Our results are more visually plausible. Briefly, instead of working in the pixel color space that can potentially cause the color fidelity problem, our method fuses the low-resolution RGB and high-resolution feature maps in the feature space and learns to fuse them into correct colors, thus alleviating the color ambiguities/artifacts at fine details. For example, In Figure 7, the wall of our results is less noisy and more accurate than the results from other methods that are either blurred or inconsistent with the ground truth. In the second example, our method produces high-frequency geometric details in the wine basket area that well differentiates the mesh color and the background color.


Table~\ref{table:comp_sbmc} reports the comparison on the Gharbi dataset~\cite{gharbi2019sample}. Following MSSPL~\cite{gharbi2019sample}, we directly test our models pretrained on the BCR dataset without fine-tuning as the training set of the Gharbi dataset is not available.  Our $\times2$ model wins 0.27dB and 0.17dB in terms of PSNR on 4spp and 8spp, respectively. When the spp is 16, our PSNR is slightly lower than MSSPL~\cite{hou2021fast}. We would like point out our method takes less high-resolution information than MSSPL. Our input high-resolution auxiliary features only include the albedo and normal, while MSSPL also takes all the shading layers as inputs. When the high resolution input is rendered at a high spp, the shading layers can contribute a lot of high frequency information. Similar to the findings in MSSPL~\cite{hou2021fast}, our results on RelMSE are heavily affected by a small number of pixels with abnormal large errors. Excluding these abnormal pixels can greatly improve our scores on RelMSE. As shown in Figure \ref{fig:comp_sbmc}, our method produces high-quality results with much fewer artifacts when compared to the ground truth.

\subsection{Discussions}


\textbf{Auxiliary features sampling rates}. As discussed above and shown in Figure~\ref{fig:aba_spp}, using more samples to generate the auxiliary features help our method generate better super resolution results. However, even using one sample per pixel to generate the auxiliary features can already enable our method to significantly outperform standard super resolution methods. Moreover, when we use 16 samples to generate these features, our results are already very close to the results that use the features generated using 4000 samples per pixel denoted as $A_{gt}$ in the figure.



\noindent \textbf{Input layers of auxiliary features}. We examine how our method works with different auxiliary feature layers. The upsampling scale is set to $4\times$. We use 4000 spp for $I_{LR}$ and $A$. As shown in Table~\ref{table:aux_layers}, both albedo and normal can improve the results significantly, as they can provide the essential high frequency visual details for super resolution. The performance of our network can be further improved if we take both of them as inputs. These findings are consistent with previous denoising methods~\cite{bako2017kernel,gharbi2019sample} where intermediate layers can improve the final results.

\noindent \textbf{Ablation study w.r.t. MSSPL~\cite{hou2021fast}}. We evaluated the performance of both our method and MSSPL~\cite{hou2021fast} using fast-to-compute auxiliary features as well as full auxiliary features. In the experiments, the upsampling scale is set to $\times4$. As shown in Table~\ref{table:aba_fast_buffer}, both our network and MSSPL benefit from using the full auxiliary features due to the richer high-resolution information they provide. However, our method with fast-to-compute layers still outperforms MSSPL with full auxiliary layers, which demonstrates the effectiveness of our network architecture.


\begin{table}[t]
    \centering

    \small
    \setlength{\tabcolsep}{10pt}
    \begin{tabular}{ccccc}
            \toprule
            Network & RDB &RSTB  &RDST &RDST + XM \\ \midrule
            PSNR   &35.56     &36.63   &37.27   &\textbf{37.45}\\
            RelMSE &0.0034    &0.0098  &0.0022  &\textbf{0.0021}\\
            \bottomrule
    \end{tabular}\vspace{-0.in}
     \caption{The effects of network architecture components on the BCR dataset~\cite{hou2021fast}. We compare the proposed RDST with RDB~\cite{zhang2018residual} and RSTB~\cite{liang2021swinir}.} \vspace{-0.1in}
     \label{table:arch}
\end{table}

\begin{table}
    \centering
    \small
    \begin{tabular}{P{1.5cm}P{0.45cm}P{0.9cm}P{0.45cm}P{0.9cm}P{0.45cm}P{0.9cm}}
            \toprule
            \multirow{2}[2]{*}{Method} & \multicolumn{2}{c}{2spp} & \multicolumn{2}{c}{4spp} & \multicolumn{2}{c}{8spp}  
            \\ \cmidrule(lr){2-3}  \cmidrule(lr){4-5}  \cmidrule(lr){6-7} 
             &PSNR &LPIPS &PSNR &LPIPS &PSNR &LPIPS 
             \\ \midrule
             AdvMC-ft &30.33   &0.209  &32.30  &0.155 &33.69 &0.126 \\
             MCSA-ft &32.68  &0.108 &34.81 & 0.080 &36.61 & 0.068  \\ \midrule
             Ours &\textbf{34.12}  &\textbf{0.090} &\textbf{35.49} &\textbf{0.070} &\textbf{37.09} &\textbf{0.057} \\
            \bottomrule
    \end{tabular}\vspace{-0.in}
        \caption{Comparison on the perceptual quality on the BCR dataset~\cite{hou2021fast}. We utilize the LPIPS~\cite{zhang2018perceptual} metric as a measure of perceptual quality.} \label{table:comp_perceptual} \vspace{-0.in}
        
\end{table}

\begin{table}
    \centering
    \small
    \begin{tabular}{P{1.5cm}P{0.9cm}P{0.9cm}P{0.9cm}P{0.9cm}P{0.9cm}}
            \toprule
             RDST Num &PSNR &RelMSE &Flops(T) &Macs(G) &Params(M) 
             \\ \midrule
             5 &34.08 &0.0046 &1.45 &723.68 &9.36 \\
             3 &33.47 &0.0056 &1.04 &519.13 &6.21 \\
             1 &32.60 &0.0091 &0.63 &314.59 &3.06 \\
            \bottomrule
    \end{tabular}\vspace{-0.in}
        \caption{The effects of the number of RDST blocks on the BCR dataset~\cite{hou2021fast}.We measured the flops and macs for a single $1024\times1024$ image~\cite{rasley2020deepspeed}. } \label{table:aba_rdst} \vspace{-0.in}
\end{table}

\begin{table}
    \centering
    \small
    \begin{tabular}{P{1.5cm}P{0.9cm}P{0.9cm}P{0.9cm}P{0.9cm}P{0.9cm}}
            \toprule
             XDG Num &PSNR &RelMSE &Flops(T) &Macs(G) &Params(M) 
             \\ \midrule
             3 &34.08 &0.0046 &1.45 &723.68 &9.36 \\
             2 &33.19 &0.0066 &1.02 &507.86 &6.42 \\
             1 &32.30 &0.1193 &0.59 &292.04 &3.49 \\
            \bottomrule
    \end{tabular}\vspace{-0.in}
        \caption{The effects of the number of XDG blocks on the BCR dataset~\cite{hou2021fast}. We measured the flops and macs for a single $1024\times1024$ image~\cite{rasley2020deepspeed}.} \label{table:aba_xdg} \vspace{-0.in}
\end{table}

\noindent \textbf{Network Effectiveness.} We examine how our network architecture works by comparing to AdvMC~\cite{xu2019adversarial} and MCSA~\cite{yu2021monte}. Specifically, we feed high-resolution 1-spp RGB and 1-spp auxiliary buffers to AdvMC and MCSA and fine tune them on the BCR dataset. In this experiment, our method takes 4-spp low-resolution RGB ($\times2$, effectively the same sampling rate as 1 spp at the high resolution) and 1-spp high-resolution auxiliary buffers. Table~\ref{table:arch_denoise} shows our method outperforms these methods, which demonstrates the effectiveness of our transformer-based network architecture.

\noindent \textbf{Network architecture components}. We examine the effect of the network architecture. The upsampling scale is set to $4\times$. In this test, we remove XM modules and replace our RDST with the state-of-the-art blocks, including RDB from RRN~\cite{zhang2018residual} and RSTB from SwinIR~\cite{liang2021swinir}. As shown in Table~\ref{table:arch}, our RDST can greatly improve the results. These improvements can be attributed to the strong generalization capability of RDST. Besides, XM modules can further improve the results. 

\noindent \textbf{Number of RDST blocks.} We examine how our network architecture works with different RDST blocks in each XDG block on the BCR dataset~\cite{hou2021fast}. In this test, the upsampling scale is set to x4. To check the impact of RDST, we set the XDG number as 3, and we investigated our results across different RDST numbers of each XDG block, including 1, 3, and 5. Besides, we also measure the flops, macs, and parameters for a single 1024x1024 image~\cite{rasley2020deepspeed}. As shown in Table~\ref{table:aba_rdst}, decreasing the number of RDST blocks accelerates the network but compromises performance.


\noindent \textbf{Number of XDG blocks.} Similar to RDST, we investigate our results across different XDG numbers, including 1, 2, and 3. The upsampling scale is set to x4 and the number of RDST of each XDG block is set to 5. As the results reported in Table~\ref{table:aba_xdg}, reducing the number of XDG blocks accelerates the network but also compromises performance. 


\noindent \textbf{Our robust loss vs SMAPE loss~\cite{meade1986long}.} Our robust loss is used based on our observations that there are a very small number of pixels with abnormally large intensity values in our dataset, mostly due to the firefly artifacts. These pixels will often incur very large errors during training and thus compromise the performance of our model. We use the robust loss to reduce these undesirable impacts of these pixels as this robust loss will limit the maximal loss value to 1 no matter how large the pixel error is. We compared these two loss functions. In our experiments, the upsampling factor is set to 4, and we set the sampling rate to (16 - 1). Models trained with the SMAPE loss showed slightly worse results: 33.96 v.s.34.12 in PSNR, and 0.0046 v.s.0.0035 in RelMSE.


        

\noindent \textbf{Super resolution scale.}
 We investigate our results across multiple scales, including $\times1$, $\times2$, $\times4$, and $\times8$. Among them, scales $\times1$ and $\times8$ exhibit weaker performance compared to $\times2$ and $\times4$. When comparing scales $\times4$ and $\times2$, $\times4$ takes less peak memory and is faster than $\times2$, but $\times2$ leads to better quality. To make a fair comparison, we maintain a consistent average sampling rate across different scales. Consequently, the low-resolution input of our $\times1$ model is rendered at a much lower average sampling rate than that of our $\times2$ model. This makes the resulting input RGB image to our model very noisy for $\times1$ and thus comprises the final quality of Ours $\times1$, as reported in the 2-spp column of Table~\ref{table:comp_bcr}. In the 4-spp column of the same table, the difference between Ours $\times1$ and Ours $\times2$ is less significant as in this setting, the average sampling rate of Ours $\times1$ is reasonably higher and provides more information for our model to synthesize higher-quality results. 

In addition, we used the same training pipeline for our $\times1$ model as we did for other scales, keeping the number of epochs consistent across all scales. However, due to the high memory requirement to train the  $\times1$ model, we have to set a smaller mini-batch size. This would also potentially impact the performance, but we believe that this is not as significant as the first reason we discussed above.

\noindent \textbf{Perceptual quality.} We examine the perceptual quality of our results using the LPIPS metric~\cite{zhang2018perceptual}. Table~\ref{table:arch_denoise} and Table~\ref{table:comp_perceptual} present the results for AdvMC~\cite{xu2019adversarial}, MCSA~\cite{yu2021monte}, and our method. Our approach outperforms the others in terms of both PSNR and LPIPS, thereby demonstrating its ability to generate images with high perceptual quality.

\vspace{0.05in}

\section{Limitations and Future Work}
\label{sec:lim}

\begin{figure}[t]
    \setlength{\indentspace}{-4.2mm}
   
    \begin{adjustbox}{valign=t}
    \tiny
    \hspace{-3mm}
        \begin{tabular}{ccc}
            \includegraphics[width=\subfigfailuresize\textwidth]{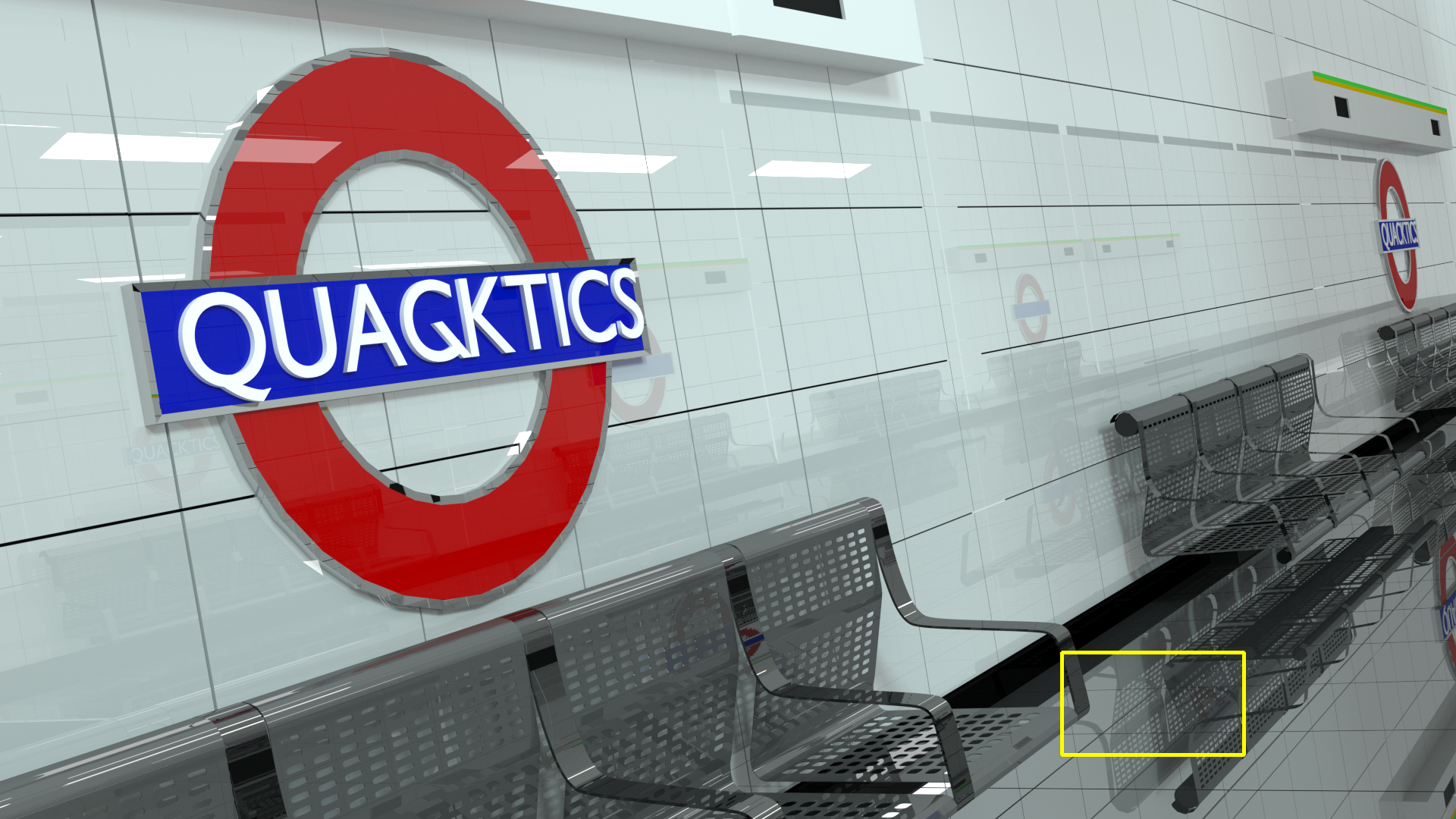} \hspace{\indentspace} &
            \includegraphics[width=\subfigfailuresize\textwidth]{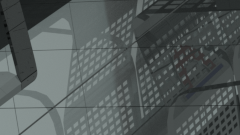} \hspace{\indentspace} &
            \includegraphics[width=\subfigfailuresize\textwidth]{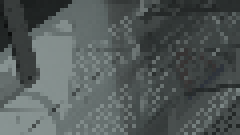}
            \\
            Source \hspace{\indentspace} &
            GT \hspace{\indentspace} &
            LR 
            \\
            \includegraphics[width=\subfigfailuresize\textwidth]{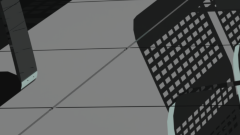} \hspace{\indentspace} &
            \includegraphics[width=\subfigfailuresize\textwidth]{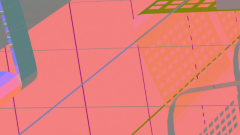} \hspace{\indentspace} &
            \includegraphics[width=\subfigfailuresize\textwidth]{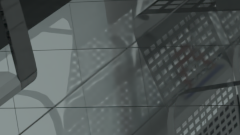} 
            \\ 
            Albedo \hspace{\indentspace} &
            Normal \hspace{\indentspace} &
            Ours
            \\
        \end{tabular}
    \end{adjustbox} 
    \caption{
        Failure example. The performance of our method is compromised in the area where the albedo and normal could not provide high-frequency details. 
    }
    \label{fig:failure_example}
    \vspace{-0.1in}
        
\end{figure}

The fusion for the high-reflection parts is challenging. Our method produces high-frequency visual details by two means: 1) train a neural network to learn to recover high-frequency information from low-resolution input and 2) use high-frequency information from the high-resolution albedo and normal maps. Our neural network can learn to produce visual details for many examples. However, super resolution from a low-resolution input alone is necessarily an ill-posed problem. In the high-reflection parts of the scene, such as the example shown in Figure~\ref{fig:failure_example}, when the high-resolution normal and albedo map cannot, by its nature, provide high-frequency details in those regions, our method may fail. 


Compared to CNN-based methods, our method is slow. However, compared to another Transformer-based method~\cite{yu2021monte}, our method uses less peak memory (0.89Gb vs 30.56Gb) and is faster (1.0s vs 2.5s) when producing a 1024x1024 image using an Nvidia A40. Research on fast transformers has been advancing quickly recently.  Patro et al. [PA23] offer an extensive review of efficient vision transformers. Through the advancement of effective token mixing strategies and efficient MLP layers, vision transformers can be significantly accelerated~\cite{li2022uniformer,guo2022cmt,yao2022wave}. For example, both CMT~\cite{guo2022cmt} and WaveViT~\cite{yao2022wave} outperform EfficientNet~\cite{tan2019efficientnet} while maintaining a lower computational complexity. Moreover, several transformer hardware accelerators have been introduced to expedite Transformer networks, such as SwiftTron~\cite{marchisio2023swifttron}. We believe that our method can benefit from the quick advance of research on Transformer.

In  this paper, we specifically explored albedo and normal as quick-to-compute auxiliary features. However, we acknowledge that other auxiliary features, such as a whitted ray-traced layer, could offer valuable high-frequency information and be generated fast. Incorporating such a layer can potentially improve the performance of our method. Unfortunately, the BCR dataset doesn’t contain such layers. We plan to explore this in our future research.

\vspace{0.05in}

\section{Conclusion}

This paper explored high-resolution fast-to-compute auxiliary features to guide super resolution of Monte Carlo renderings. We developed a dedicated cross-modality Transformer network to fuse high-resolution fast-to-compute auxiliary features with the corresponding low-resolution rendering. We designed a Transformer-based cross-modality module to fuse the features from two modalities. We also developed a Residual Densely-connected Swin Transformer block to learn more representative features. Experimental results indicate that our proposed method surpasses existing state-of-the-art super-resolution and denoising techniques in producing high-quality images.



\printbibliography   

\end{document}